\documentclass{jfm}
\usepackage{graphicx}
\usepackage{epstopdf, epsfig}
\usepackage{color}
\usepackage{graphicx}

\shorttitle{Wave coupling in liquid metal batteries}
\shortauthor{G. M. Horstmann, N. Weber and T. Weier}

\title{Coupling and stability of interfacial waves in liquid metal batteries}

\author{G. M. Horstmann\aff{1}
  \corresp{\email{g.horstmann@hzdr.de}},
  N. Weber\aff{1}
 \and T. Weier\aff{1}}

\affiliation{\aff{1}Institute of Fluid Dynamics, Helmholtz-Zentrum Dresden - Rossendorf, Bautzner Landstr. 400, 01328 Dresden, Germany}

\begin{document}

\maketitle

\begin{abstract}
We investigate the interfacial wave coupling dynamics in liquid metal batteries and their effects to the battery's operation safety. Similar to aluminum reduction cells, liquid metal batteries can be highly susceptible to magnetohydrodynamical instabilities that excite undesired interfacial waves capable to provoke short-circuits. However, in liquid metal batteries the wave dynamics is far more complex since two metal-electrolyte interfaces are present that may step into resonance. In the first part of this paper, we present a Potential analysis of coupled gravity-capillary interfacial waves in a three-layer battery model of cylindrical shape. Analytical expressions for the amplitude ratio and the wave frequencies are derived and it is shown that the wave coupling can be completely described by two independent dimensionless parameters. We provide a decoupling criterion clarifying that wave coupling will be present in most future liquid metal batteries. In the second part, the theory is validated by comparing it with multiphase direct numerical simulations. An accompanying parameter study is conducted to analyze the system stability for differently strongly coupled interfaces. Three different coupling regimes are identified involving characteristic coupling dynamics. For strongly coupled interfaces we observe novel instabilities that may have beneficial effects on the operational safety. 
\end{abstract}

\begin{keywords}

\end{keywords}

\section{Introduction}
\label{sec:Introduction}
Liquid metal batteries (LMBs) are discussed today as a cheap grid scale energy storage, as required for the deployment of fluctuating renewable energies. The basic operation principle of LMBs is very simple. They consist of three immiscible liquid layers, filled in a closed insulating container, that self-stratify above each other on the basis of their differences in density. A light alkaline metal floats on the top, a heavy metal or half-metal alloy is placed at the bottom, and a thin layer of a molten-salt electrolyte is sandwiched in between. The electrolyte is chosen to be conductive to positive ions of the light metal. \\
The liquid-liquid salt-metal interfaces facilitate super-fast charge transfer and high current densities making the LMB, in combination with low Ohmic losses, an efficient and flexible stationary energy storage \citep{Kim2013b}. Further, LMBs offer high potential of being low-cost due to the simple fabrication and size scalability in comparison to conventional batteries as well as long life cycles naturally favoured by the liquid nature preventing microstructural degradation mechanisms. \\
However, the easy scalability to large-size batteries together with the very high cell current densities that can reach up to $13\, {\rm A}/{\rm cm}^2$ \citep{Cairns1969b} can lead to instability problems interesting for fluid dynamicists. Large-size cells are potentially susceptible to magnetohydrodynamical (MHD) instabilities that may have positive effects on the one hand, e.g. fluid motion in the metal layers can improve the mixing of reactants. On the other hand, strong metal flows can severely deform the interfaces such that the salt layer may be disrupted causing a short-circuit. Hence, it is of crucial importance to study the impacts of flow instabilities on the batteries to enable a secure operation in future large cells. \\
Several instability mechanism have been identified in the last decade that can arise in LMBs. Primarily, the Tayler instability (TI) has been intensively studied \citep{Stefani2011,Weber2013,Weber2015b,Herreman2015} and it was found that the TI can be reliably controlled e.g. by applying axial magnetic fields, central counter currents or modifying the cell geometry. Further, electro-vortex flows \citep{Weber2014b,Stefani2015} as well as thermal and Marangoni convection due to volumetric Joule heating in the electrolyte \citep{Kelley2014,Shen2015,Koellner2017} have been investigated. However, in recent years long-wave interfacial instabilities, in particular the so-called metal pad roll instability (MPRI), emerged as the possibly most critical mechanism regarding the operation safety of shallow batteries \citep{Weber2017}. An extensive overview of MHD instabilities investigated in LMBs is given by \cite{Weier2017}.\\
\begin{figure}             % Einbetten in figure wie gehabt
	\centering                  % zentrierte Ausrichtung, optional
	\def\svgwidth{250pt}    
	% die Bildbreite muss auf diese Weise festgelegt werden!
	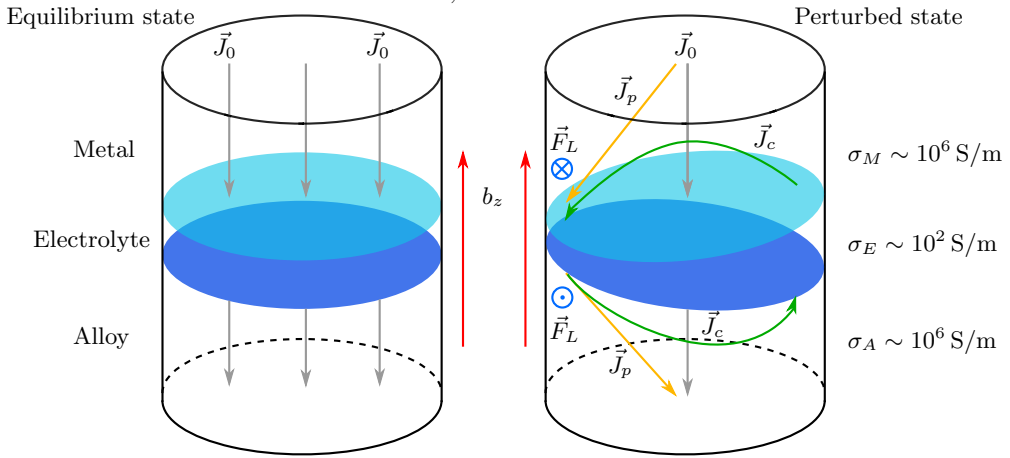  
	% 'versuchsaufbau' durch Dateinamen ersetzen.
	\caption{Schematic explanation of the MPRI in LMBs. If both interfaces are at rest (left cell), the cell current density $\vec{J}_0$ is purely vertical for idealized boundary conditions and does not interact with likewise vertical external magnetic fields $b_z$. Once one or both interfaces are somehow displaced, non-vertical perturbation currents $\vec{J}_p$ arise that are deflected to the left wall where the low conducting electrolyte layer is thin. $\vec{J}_p$ comprises a horizontal component, the compensation current $\vec{J}_c$, that induces horizontal Lorentz forces $\vec{F}_L$ in interaction with $b_z$, finally driving a rotational wave motion.}    % Bildunterschrift, optional
	\label{fig:Sele}          % Label für Verweise, optional
\end{figure}
In the paper at hand, we aim to give a deeper understanding how the MPRI is manifested in LMBs incorporating hydrodynamically coupled interfaces. Originally, the MPRI was discovered in Hall-H\'{e}roult aluminum reduction cells (ARCs) that are physically similar to LMBs. ARCs consist of two liquid layers, molten aluminum at the bottom and a cryolite bath with dissolved aluminum floating on the top, that is electrochemically reduced by applying strong vertical cell currents. \cite{Sele1977} was the first to give a qualitative explanation of the  MPRI for two-layer reduction cells. Figure \ref{fig:Sele} illustrates his fundamental ideas expanded to three-layer LMBs: the key mechanism driving the metal pad is traced back to the interaction of horizontal currents with the vertical component of an external magnetic field $\vec{B}$ that can be induced, e.g, by supply lines. For simplicity, let us assume that the cell current density $\vec{J}_0$ is purely vertical during the charging or discharging and the magnetic field purely vertical and homogeneous $\vec{B}=b_z \vec{e}_z$. For fluid layers at rest, $\vec{J}_0$ and $b_z$ are parallel such that no motion is expected to arise in this idealized configuration (left cell). However, once one or both interfaces are somehow displaced (right cell), the current will be deflected. That is due to the orders of magnitude lower electrical conductivity of the electrolyte $\sigma_E$ in comparison to the electrode metals. The current must always take the way of least resistance that means here where the electrolyte layer is thin. As a result, a redistributed non-vertical perturbation current $\vec{J}_p$ arises that can be decomposed into the unperturbed vertical current $\vec{J}_0$ and a horizontal component that is denoted as the compensation current $\vec{J}_c$. Together with $b_z$ the latter horizontal currents lead to horizontal Lorentz forces $\vec{F}_L$ here pointing inwards at the upper light metal and outwards at the heavy one. Finally, these Lorentz forces, which are perpendicular to the interfacial slope, drive the metals leading to rotational interfacial wave motion that is called the \textit{metal pad roll}. Though, in contrast to the ARCs, from this picture it is not clear in which direction both interfaces will rotate since the induced Lorentz forces are oppositely directed. We leave that question open to section \ref{sec:Weakly}. \\
While the scenario sketched above provides a first intuitive feeling about the origin of the MPRI, the physical nature is actually far more complex. Stability analysis were performed in the last decades by \cite{Sneyd1994,Bojarevics1994,Davidson1998} showing that the Lorentz force can couple standing gravity waves leading to a rotation and exponential growth in the unstable limit. The instability mechanism can be described just as well by wave reflection at the tank walls \citep{Lukyanov2001}. \\
Nevertheless, far less is known about the MPRI in LMBs yet. \cite{Zikanov2015} first investigated long-wave coupling and stability using a simplified mechanical analogy inspired by \cite{Davidson2001}. Very recently, the MPRI was numerically found in LMB models and identified as potentially dangerous using 2D shallow water \citep{Bojarevics2017,Zikanov2017} and 3D direct numerical simulations \citep{Weber2017}. Both studies examined magnesium-antimony (Mg$||$Sb) batteries, where essentially only the upper interface was found to be excited. For this particular case LMBs act very similarly to ARCs. \cite{Weber2017} found that the stability can be roughly predicted by one dimensionless parameter $\beta$, originally proposed by \cite{Sele1977} for ARCs, reading
\begin{equation}
\beta = \frac{I_0 b_z}{h_E h_M (\rho_E -\rho_M)g} \label{eq:Sele}
\end{equation}
with $I_0$ denoting the total cell current, $h_E$ and $h_M$ the initial heights of the electrolyte and light metal layers, $\rho_E$ and $\rho_M$ the densities of the electrolyte and the light metal, and $g$ the standard acceleration due to gravity. $\beta$ can be considered as the ratio of the magnetic force arising due to some interface displacement to the gravity force exerted on the interface for the same displacement. The metal pad roll is predicted to arise after exceeding some critical value $\beta > \beta_{\rm c}$ that generally depends on the viscosity, interfacial tension and the cell geometry. \\
However, this criterion can not be generally applied to LMBs. From  (\ref{eq:Sele}) it is clear, that small density differences $\rho_E - \rho_M$ are destabilizing. In the investigated Mg$||$Sb cells, the density difference at the upper interface is much lower than the density difference at the lower one ($\rho_E - \rho_M << \rho_A - \rho_E$) clarifying why only the upper interface is excited. Actually, that is not the case for most possible electrode metal combinations as we will show in section \ref{sec:CouplingCriterion}. On that basis, this paper investigates how the MPRI is manifested in universal LMBs with arbitrary electrode metals where both interfaces may be excited even simultaneously. \\
In the first part, we present a theoretical Potential analysis describing the hydrodynamical coupling of gravity-capillary waves bounded in cylindrical cells. Two independent dimensionless parameters were derived which completely determine the wave coupling. We found further that the interfaces can be coupled by two different coupling modes. In the second part, we compare the theory with direct numerical simulations (DNS). A parameter study was performed to investigate the impact of coupled interfaces on the stability. Three different coupling regimes were identified involving different characteristic coupling states and different stability properties. For strongly coupled LMBs we found novel types of instabilities that can not be described by the classical MPRI mechanism alone.
\section{Theoretical approach}
For a better understanding of the wave dynamics in LMBs, it is at first of importance to study the coupling behavior of the two electrolyte-metal interfaces. The essential difference to aluminum reduction cells is manifested in the presence of the second interface that may influence both the overall liquid motion as well as the stability of the LMB. Once one of the interfaces is somehow excited e.g by the induced Lorentz forces, the second interface can be also affected since the interfaces can be both electromagnetically and hydrodynamically coupled. On the one hand, the current distribution can be changed by interfacial displacements that may lead to destabilizing horizontal currents also near the second interface. On the other hand, both interfaces can mutually exert pressure forces. So far, very little about these coupling dynamics is known in literature. It has been investigated only by \cite{Zikanov2015} using a simplified mechanical analogy that is not sufficient to reflect the dispersion relations of continuous interfacial fluid waves and very recently by \cite{Zikanov2017} using the St. Venant shallow water model. Hence, a more detailed MHD analysis of three-layer systems is necessary to get deeper insights into the sloshing dynamics of LMBs. \\
In the following analysis we only focus on the first point, the pressure coupling, and completely neglect all electromagnetic effects as a first approach since a full magnetohydrodynamical description is very costly. However, this approach does not restrict the applicability too much, because it is well known from literature that the MPRI causes in good approximation traveling natural gravity waves in aluminum reduction cells \citep{Bojarevics1994,Gerbeau2006,Molokov2011}. Recently, this was found to hold true also in a three-layer LMB model \citep{Weber2017}. \\
Based on these findings, we applied Potential theory to describe coupled interfacial gravity-capillary waves in bounded three-layer systems without using the commonly applied shallow water approximation. We are quite aware that capillary effects are of little importance in future large-size LMBs, however, on lab-scale both our numerical and planned experimental LMB models have small diameters of the order $D \sim 10^{-1}\, {\rm m}$, where interfacial tension indeed influences the wave frequencies. In order to allow sufficient comparisons with our experiments, we have decided to take into account interfacial tensions. \\
Although Potential theory has been available for already more than 200 years, theoretical descriptions of coupled interfacial waves in multilayer systems are very rare in literature. There are only a few recent publications in the context of oceanography \citep{Woolfenden2011,Mohapatra2011,Issenmann2016} describing the coupling of an infinite internal wave with an infinite free-surface in a one-dimensional two-layer system. However, these approaches are still insufficient to describe interfacial wave motion in LMBs demanding a three-dimensional analysis of a bounded three-layer system considering both gravity and capillary forces at both interfaces. By incorporating these extensions the coupling dynamic becomes far more complex in comparison to the oceanographic systems, and new physical effects appear as we will show in the following sections.
\subsection{Mathematical setting}
\begin{figure}             % Einbetten in figure wie gehabt
	\centering                  % zentrierte Ausrichtung, optional
	\def\svgwidth{180pt}    
	% die Bildbreite muss auf diese Weise festgelegt werden!
	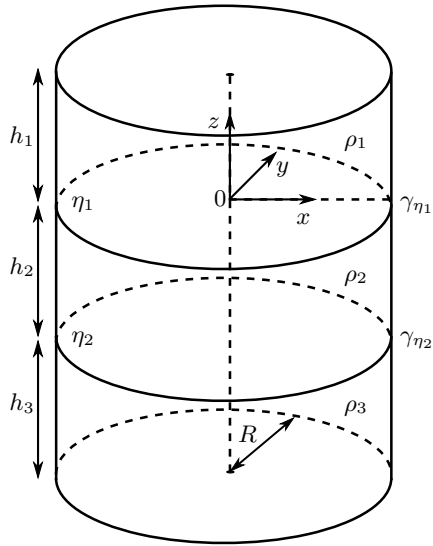  
	% 'versuchsaufbau' durch Dateinamen ersetzen.
	\caption{Sketch of the idealized cylindrical three-layer system determined by the radius $R$, three phases $i = 1,2,3$ of densities $\rho_i$ and heights $h_i$ as well as two interfaces $\eta_1$ and $\eta_2$ with corresponding interfacial tensions $\gamma_{\eta_1}$ and $\gamma_{\eta_2}$.}    % Bildunterschrift, optional
	\label{fig:Three_Layer}          % Label für Verweise, optional
\end{figure}
We apply the Potential analysis to an ideal cylinder, as depicted in figure \ref{fig:Three_Layer}. The cylindrical geometry was chosen because it is the most unstable geometry, where the cell is always unstable ($\beta_c = 0$) for inviscid flows \citep{Lukyanov2001}. Since we intended to analyze the relevance of the interfacial coupling dynamics to the global instability, the cylinder was our generic geometry of choice to exclude the contribution of the horizontal aspect ratio to the stability. Anyway, the theory can be easily transfered to arbitrary rectangular boxes and provides qualitatively the same results. \\
The cylindrical tank with the radius $R$ was defined to contain three immiscible phases $i=1,2,3$ of densities $\rho_i$ and heights $h_i$, where $\rho_1 < \rho_2 < \rho_3$ has to be fulfilled to realize a self-stratified system. Two interfaces with interfacial tensions $\gamma_{\eta_1}$ and $\gamma_{\eta_2}$ are located at $z=\eta_1(x,y,t)$ and $z=\eta_2(x,y,t)$. The coordinate origin of the system is placed in the center of the upper unperturbed interface $\eta_1$. \\
To facilitate Potential theory, a few simplifications have to be supposed. At first, an inviscid flow has to be assumed, by which damping effects at the tank walls and on the interfaces are neglected. Actually, viscosity affects the natural frequencies of sloshing interfaces only very slightly \citep{Ibrahim2005} and can be unproblematically neglected in comparison to the pressure exchange forces basically determining the coupling dynamics. Secondly, the flow is supposed to be irrotational, which at the first glance seems to be an counterintuitive assumption since the MPRI is driven by rotational Lorentz forces. Nevertheless, as mentioned before, in very good approximation natural irrotational gravity modes are excited. In section \ref{sec:Weakly} it is shown that rotational Lorentz forces primarily influence the shape of the waves, but not the frequencies and amplitude ratios. At last, the wave amplitudes are restricted to be sufficiently small to facilitate linear wave theory and all three fluids are assumed to be incompressible.
\subsection{Potential analysis}
\begin{figure}           % Einbetten in figure wie gehabt
	\centering                  % zentrierte Ausrichtung, optional
	\def\svgwidth{300pt}    
	% die Bildbreite muss auf diese Weise festgelegt werden!
	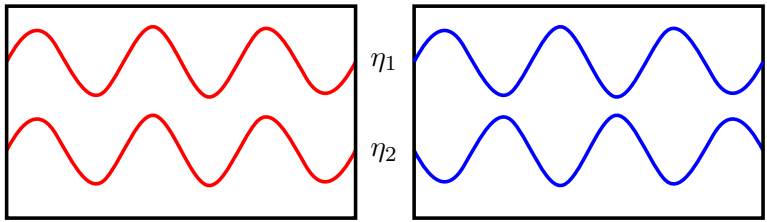  
	% 'versuchsaufbau' durch Dateinamen ersetzen.
	\caption{Schematic sketch of the symmetric fast mode $\omega_+$ (red) and the antisymmetric slow mode $\omega_-$ (blue).}    % Bildunterschrift, optional
	\label{fig:Modes}          % Label für Verweise, optional
\end{figure}

Claiming the above simplifications the velocity fields $\vec{u}_i$ of all three phases $i=1,2,3$ can be expressed unambiguously by the gradient of  scalar potentials $\phi_i$
\begin{equation}
\vec{u}_i = \vec{\nabla}\phi_i.
\label{eq:Pot}
\end{equation}
For fulfilling momentum and mass conservation, Bernoulli's and Laplace's equation have to be solved in all phases
\begin{eqnarray}
&&\frac{\partial \phi_i}{\partial t} + \frac{P_i}{\rho_i} + gz = {\rm const} \label{eq:Bernoulli} \\
&&\Delta \phi_i = 0 \label{eq:Laplace},
\end{eqnarray} 
where $P_i$ denotes the pressure and $g$ the standard acceleration due to gravity. In addition, using cylindrical coordinates ($r,\varphi ,z$), the following sets of dynamic boundary conditions must be satisfied:
\begin{eqnarray}
{\rm Top \ wall{:}}& \ \ \ &\frac{\partial \phi_1}{\partial z} = 0|_{z = h_1} \label{2a} \\
{\rm Bottom \ wall{:}}& \ \ \ &\frac{\partial \phi_3}{\partial z} = 0|_{z = - (h_2 + h_3)}\label{2c}\\
{\rm Side \ wall{:}}& \ \ \ &\frac{\partial \phi_1}{\partial r} = \frac{\partial \phi_2}{\partial r} =\frac{\partial \phi_3}{\partial r} = 0|_{r = R} \label{1c}\\
{\rm Upper \ interface{:}}& \ \ \ &\frac{\partial \phi_1}{\partial z} = \frac{\partial \phi_2}{\partial z}|_{z=\eta_1}\label{3a}\\
{\rm Lower \ interface{:}}& \ \ \ &\frac{\partial \phi_2}{\partial z} = \frac{\partial \phi_3}{\partial z}|_{z=\eta_2}\label{3b}
\end{eqnarray}
No-outflow boundary conditions were formulated at the side wall $r=R$ (\ref{1c}), at the top $z=h_1$ (\ref{2a}) and the bottom wall $z=-(h_2 + h_3)$ (\ref{2c}), while conditions (\ref{3a}) and (\ref{3b}) ensure interface preservation. By assuming a harmonic time dependency $ \phi_i \sim \cos(\omega t)$, where $\omega$ marks the angular frequency, Potential solution for the Laplace equations (\ref{eq:Laplace}) in cylindrical coordinates fulfilling all above conditions can be found as superpositions of infinite wave modes $m \in \mathbb{N}_0$  and $n \in \mathbb{N}_1$:
\begin{subeqnarray}
\phi_{1}(r,\varphi,z,t) &=& - \sum_{m=0}^{\infty}\sum_{n=1}^{\infty}\tilde{\eta}_{1}^{mn}\frac{\cosh(\frac{\epsilon_{mn}}{R}(z-h_1))}{\sinh(\frac{\epsilon_{mn}}{R} h_1))}J_m (\epsilon_{mn}\frac{r}{R})\cos(m\varphi - \omega t) \label{eq:Pot1} \\
\phi_{2}(r,\varphi,z,t) &=& \sum_{m=0}^{\infty}\sum_{n=1}^{\infty} 
\left [\tilde{\eta}_{1}^{mn} \left( e^{\frac{\epsilon_{mn}}{R} h_2}\cosh(\frac{\epsilon_{mn}}{R} z) - e^{-\frac{\epsilon_{mn}}{R} z}\sinh(\frac{\epsilon_{mn}}{R} h_2) \right) \right. \nonumber \\
&&\left. - \tilde{\eta}_{2}^{mn} \cosh(\frac{\epsilon_{mn}}{R} z) \right] \cdot \frac{1}{\sinh(\frac{\epsilon_{mn}}{R} h_2)} J_m (\epsilon_{mn}\frac{r}{R})\cos(m\varphi - \omega t) \\
\phi_{3}(r,\varphi,z,t) &=& \sum_{m=0}^{\infty}\sum_{n=1}^{\infty}\tilde{\eta}_{2}^{mn}\frac{\cosh(\frac{\epsilon_{mn}}{R}(z+ h_2 + h_3))}{\sinh(\frac{\epsilon_{mn}}{R} h_3))}J_m (\epsilon_{mn}\frac{r}{R})\cos(m\varphi - \omega t). \label{eq:Pot3}
\end{subeqnarray}
Here $J_m$ denotes the $m$-th order Bessel function of the first kind and $\tilde{\eta}_{1}^{mn}$ as well as $\tilde{\eta}_{2}^{mn}$ the mode-dependent amplitudes of the upper and lower interface. $m$ is often called the azimuthal wave number because it only appears in the cosine terms and for $m=0$ all solutions are radial-symmetric. Complementary, $\epsilon_{mn}$ mark the radial wave numbers, which are restricted to the $n$ roots of the first derivative of the $m$th-order Bessel function
\begin{equation}
J^{'}_m (\epsilon_{mn}) = 0,
\end{equation} 
to fulfill the radial no-outflow boundary condition and can be easily determined numerically or found in \cite{AbramowitzStegun1972}. Exploiting that the vertical velocity on the interfaces equates with their linearized time variation
\begin{equation}
\left. \frac{\partial \eta_1}{\partial t} = \frac{\partial \phi_1}{\partial z} = \frac{\partial \phi_2}{\partial z}\right|_{z = 0} \ \ {\rm and} \ \ \left. \frac{\partial \eta_2}{\partial t} = \frac{\partial \phi_2}{\partial z} = \frac{\partial \phi_3}{\partial z}\right|_{z = -h_2} \label{eq:Prev}
\end{equation}
the interfacial shapes can be easily derived from the scalar Potentials:
\begin{subeqnarray}
\eta_{1}(r,\varphi ,t) &=& \sum_{m=0}^{\infty}\sum_{n=1}^{\infty}\tilde{\eta}_{1}^{mn}\frac{\epsilon_{mn}}{R\omega}J_m (\epsilon_{mn}\frac{r}{R})\cos(m\varphi - \omega t) \label{eq:eta1} \\
\eta_{2}(r,\varphi ,t) &=& \sum_{m=0}^{\infty}\sum_{n=1}^{\infty}\tilde{\eta}_{2}^{mn}\frac{\epsilon_{mn}}{R\omega}J_m (\epsilon_{mn}\frac{r}{R})\cos(m\varphi - \omega t). \label{eq:eta2}
\end{subeqnarray}
To approximate the shape of the MPRI it is sufficient to only consider the first modes since higher modes are damped by mechanical dissipation forces like shear stresses and surface tension. It is known from aluminum reduction cells \citep{Davidson1998,Gerbeau2001} and also confirmed for the Mg$||$Sb LMB \citep{Weber2017} that in many cases the $m=n=1$ mode giving $\epsilon_{11} \approx 1.841$ becomes most unstable by the MPRI. This mode corresponds to the rolling tilted interface sketched in figure \ref{fig:Sele}.\\
Now, the wave frequencies can be determined by regarding the dynamic boundary condition, which claims that the pressure drop $\Delta P$ at the interfaces has to be balanced by the surface tensions $\gamma_{\eta_1}$ and $\gamma_{\eta_2}$. The pressure of each phase is determined by the Bernoulli equation (\ref{eq:Bernoulli}), whereby the pressure drops can be expressed with
\begin{subeqnarray}
\Delta P|_{\eta_1} &=& P_2 |_{\eta_1} - P_1 |_{\eta_1} = -\rho_2 \frac{\partial \phi_2}{\partial t}|_{\eta_1} -\rho_2 g \eta_1 + \rho_1 \frac{\partial \phi_1}{\partial t}|_{\eta_1} + \rho_1 g \eta_1 \label{eq:DynBC1}\\
\Delta P|_{\eta_2} &=& P_3 |_{\eta_2} - P_2 |_{\eta_2} = -\rho_3 \frac{\partial \phi_3}{\partial t}|_{\eta_2} -\rho_3 g \eta_2 + \rho_2 \frac{\partial \phi_2}{\partial t}|_{\eta_2} + \rho_2 g \eta_2 . \label{eq:DynBC2}
\end{subeqnarray}
The pressure drops due to surface tension are described in dependence on the principle radii of curvature $R_1$ and $R_2$ by the Young-Laplace equation\begin{subeqnarray}
\Delta P|_{\eta_1} &=& \gamma_{\eta_1}\left(\frac{1}{R_1} + \frac{1}{R_2}  \right)|_{\eta_1} \label{eq:Surf1} \\
\Delta P|_{\eta_2} &=& \gamma_{\eta_2}\left(\frac{1}{R_1} + \frac{1}{R_2}  \right)|_{\eta_2}. \label{eq:Surf2}
\end{subeqnarray} 
For small amplitudes the radii of curvature can be approximated by the second derivatives of the interfaces,
\begin{equation}
\frac{1}{R_1}|_{\eta_j} \approx \frac{\partial^2 \eta_j}{\partial x^2}, \ \ \ \frac{1}{R_2}|_{\eta_j} \approx \frac{\partial^2 \eta_j}{\partial y^2} \ \ {\rm with} \ \ j = 1,2.
\end{equation} 
By now, the dynamic boundary condition can be expressed in dependence on the flow potentials $\phi_i$ by calculating the first time derivatives of  (\ref{eq:DynBC1}), exploiting the preservations conditions (\ref{eq:Prev})
and eliminating the pressure drops. One gets
\begin{subeqnarray}
&&\left.  \rho_1 \frac{\partial^2 \phi_1}{\partial t^2} + \rho_1 g \frac{\partial \phi_1}{\partial z} =  \rho_2 \frac{\partial^2 \phi_2}{\partial t^2} + \rho_2 g \frac{\partial \phi_2}{\partial z} - \gamma_{\eta_1}\left(\frac{\partial^2}{\partial x^2} + \frac{\partial^2}{\partial y^2}  \right) \frac{\partial \phi_2}{\partial z} \right|_{z = \eta_1} \\
&&\left.  \rho_2 \frac{\partial^2 \phi_2}{\partial t^2} + \rho_2 g \frac{\partial \phi_2}{\partial z} =  \rho_3 \frac{\partial^2 \phi_3}{\partial t^2} + \rho_3 g \frac{\partial \phi_3}{\partial z} - \gamma_{\eta_2}\left(\frac{\partial^2}{\partial x^2} + \frac{\partial^2}{\partial y^2}  \right) \frac{\partial \phi_3}{\partial z} \right|_{z = \eta_2}.
\end{subeqnarray}
The interfacial tension terms can be further simplified using the Cartesian Laplace equation
\begin{equation}
\frac{\partial^2 \phi_i}{\partial z^2} = -\left(\frac{\partial^2 \phi_i}{\partial x^2} + \frac{\partial^2 \phi_i}{\partial y^2} \right)
\end{equation}
so that the kinematic boundary conditions finally read
\begin{subeqnarray}
&&\left.  \rho_1 \frac{\partial^2 \phi_1}{\partial t^2} + \rho_1 g \frac{\partial \phi_1}{\partial z} =  \rho_2 \frac{\partial^2 \phi_2}{\partial t^2} + \rho_2 g \frac{\partial \phi_2}{\partial z} + \gamma_{\eta_1}\frac{\partial^3 \phi_2}{\partial z^3} \right|_{z = \eta_1} \label{eq:Kin1} \\
&&\left.  \rho_2 \frac{\partial^2 \phi_2}{\partial t^2} + \rho_2 g \frac{\partial \phi_2}{\partial z} =  \rho_3 \frac{\partial^2 \phi_3}{\partial t^2} + \rho_3 g \frac{\partial \phi_3}{\partial z} + \gamma_{\eta_2}\frac{\partial^3 \phi_3}{\partial z^3} \right|_{z = \eta_2}. \label{eq:Kin2}
\end{subeqnarray} 
In this form the conditions likewise can be used in cylindrical coordinates since the $x$- and $y$-coordinate dependency has been eliminated. Finally, two dispersion relations $\omega(\epsilon_{mn})$ connected to both interfaces are derived by inserting the potential solutions (\ref{eq:Pot1}) into (\ref{eq:Kin1}):
\begin{subeqnarray}
\omega_{1|2mn}^{2} &=& \frac{(\rho_2 - \rho_1)g\frac{\epsilon_{mn}}{R} + \gamma_{\eta_1}\left(\frac{\epsilon_{mn}}{R} \right)^3}{\rho_1\coth(\frac{\epsilon_{mn}}{R}h_1) + \rho_2\left(\coth(\frac{\epsilon_{mn}}{R}h_2) - \frac{\tilde{\eta}_{2}^{mn}}{\tilde{\eta}_{1}^{mn}}\frac{1}{\sinh(\frac{\epsilon_{mn}}{R}h_2)}\right)} \label{eq:Omega12} \\
\omega_{2|3mn}^{2} &=& \frac{(\rho_3 - \rho_2)g\frac{\epsilon_{mn}}{R}
	+ \gamma_{\eta_2}\left(\frac{\epsilon_{mn}}{R} \right)^3}{\rho_3\coth(\frac{\epsilon_{mn}}{R}h_3) + \rho_2\left(\coth(\frac{\epsilon_{mn}}{R}h_2) - \frac{\tilde{\eta}_{1}^{mn}}{\tilde{\eta}_{2}^{mn}}\frac{1}{\sinh(\frac{\epsilon_{mn}}{R}h_2)}\right)} . \label{eq:Omega23}
\end{subeqnarray}
Both relations depend on apriori unknown amplitude ratios $\frac{\tilde{\eta}_{1}^{mn}}{\tilde{\eta}_{2}^{mn}}$ and $\frac{\tilde{\eta}_{2}^{mn}}{\tilde{\eta}_{1}^{mn}}$. 
Eliminating them leads to a fourth order dispersion relation
\begin{eqnarray}
&&a\omega^4 + b\omega^2 + c = 0, \ {\rm with} \\ 
&&a = \left(\rho_2 \coth(\frac{\epsilon_{mn}}{R}h_2)+\rho_3 \coth(\frac{\epsilon_{mn}}{R}h_3)\right)\left(\rho_2 \coth(\frac{\epsilon_{mn}}{R}h_2)+\rho_1 \coth(\frac{\epsilon_{mn}}{R}h_1)\right) \nonumber \\
&&\ \ \, -\frac{\rho_{2}^{2}}{\sinh(\frac{\epsilon_{mn}}{R}h_2)^2} \nonumber \\
&&b = -\left((\rho_2 -\rho_1)g \frac{\epsilon_{mn}}{R} + \gamma_{\eta_1}\left(\frac{\epsilon_{mn}}{R} \right)^3 \right)\left(\rho_2 \coth(\frac{\epsilon_{mn}}{R}h_2)+\rho_3 \coth(\frac{\epsilon_{mn}}{R}h_3)\right) \nonumber \\
&&\ \ \, - \left((\rho_3 - \rho_2)g \frac{\epsilon_{mn}}{R} + \gamma_{\eta_2}\left(\frac{\epsilon_{mn}}{R} \right)^3\right)\left(\rho_2 \coth(\frac{\epsilon_{mn}}{R}h_2)+\rho_1 \coth(\frac{\epsilon_{mn}}{R}h_1)\right) \nonumber \\
&&c = \left((\rho_2 -\rho_1)g \frac{\epsilon_{mn}}{R} +  \gamma_{\eta_1}\left(\frac{\epsilon_{mn}}{R} \right)^3 \right) \left((\rho_3 - \rho_2)g \frac{\epsilon_{mn}}{R} +   \gamma_{\eta_2}\left(\frac{\epsilon_{mn}}{R} \right)^3\right), \nonumber
\end{eqnarray}
which provides two independent analytical solutions
\begin{equation}
\omega_{\pm}^{2} = \frac{-b \pm \sqrt{b^2 - 4ac}}{2a}. \label{eq:SolDisp}
\end{equation}
Consequently, in contrast to two-layer ARCs, LMBs involve always two eigenfrequencies $\omega_+$ and $\omega_-$, where it can be seen readily from  (\ref{eq:SolDisp}) that $\omega_+$ is always larger than  $\omega_-$. Hence, in analogy to the coupled harmonic oscillator,  $\omega_+$ is called the fast mode and $\omega_-$ the slow mode, respectively. \\
Moreover, by knowing the frequency solutions from  (\ref{eq:SolDisp}), the amplitude ratios can be analytically determined by transposing  (\ref{eq:Omega12}$a$) or (\ref{eq:Omega23}$b$) giving
\begin{eqnarray}
\frac{\tilde{\eta}_{1}^{mn}}{\tilde{\eta}_{2}^{mn}}_{\pm} &=& \frac{\sinh(\frac{\epsilon_{mn}}{R}h_2)}{\rho_2} \Bigg[ \rho_2 \coth(\frac{\epsilon_{mn}}{R}h_2) + \rho_3 \coth(\frac{\epsilon_{mn}}{R}h_3) \Bigg. \nonumber \\ 
\Bigg. &-&\frac{(\rho_3 - \rho_2)g \frac{\epsilon_{mn}}{R} + \gamma_{\eta_2}\left(\frac{\epsilon_{mn}}{R} \right)^3}{\omega_{\pm}^2}    \Bigg]. \label{eq:Ratio}
\end{eqnarray}
Corresponding to the two frequency solutions, we also find two different amplitude ratios, whereby it can readily be shown from  (\ref{eq:Ratio}) that for the plus mode the amplitude ratio is alway positive and for the minus mode always negative
\begin{equation}
\frac{\tilde{\eta}_{1}^{mn}}{\tilde{\eta}_{2}^{mn}}_{+} > 0 , \ \ \ \ \ \ \frac{\tilde{\eta}_{1}^{mn}}{\tilde{\eta}_{2}^{mn}}_{-} < 0.
\end{equation}
From that directly follows that in the plus mode the interfaces propagate symmetrically in phase and in the minus mode antisymmetrically with a 180 degree phase shift. Exemplarily, both coupling states are schematically sketched in figure \ref{fig:Modes}. \\ 
To conclude, we have to distinguish between two essential coupling modes, the one fast and symmetric and the other slow and antisymmetric, that may both occur in LMBs having different consequences for the operation stability. This result is well known from the coupled harmonic oscillator and was to be reproduced since we employed linear wave theory. However, the formulas (\ref{eq:SolDisp}) and (\ref{eq:Ratio}) now allow us to study the natural frequencies and the coupling dynamics in dependence on nine geometrical cell parameters and material properties  $\rho_1 , \rho_2, \rho_3 , \gamma_{\eta_1}, \gamma_{\eta_2}, h_1 , h_2 , h_3 , R$ as presented in the next section.  
\subsection{Analysis of the coupling dynamics}
\begin{figure} 
	\vspace*{0.3cm}            % Einbetten in figure wie gehabt
	\centering                  % zentrierte Ausrichtung, optional
	\def\svgwidth{380pt}    
	% die Bildbreite muss auf diese Weise festgelegt werden!
	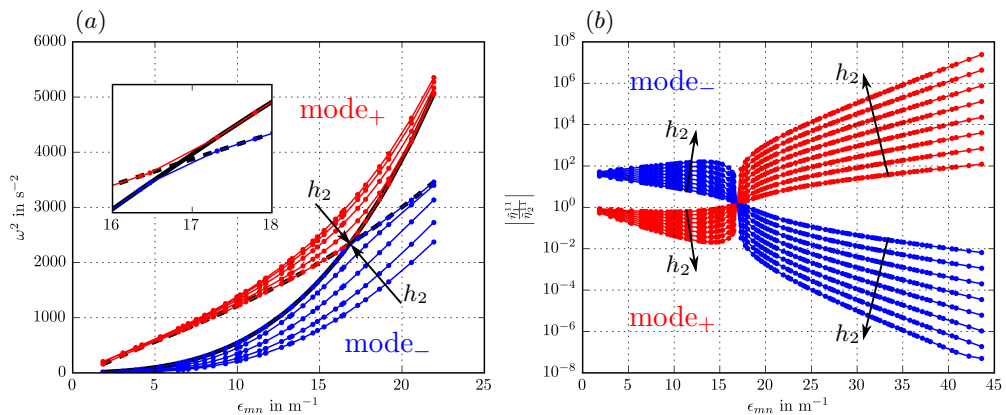  
	% 'versuchsaufbau' durch Dateinamen ersetzen.
	\caption{($a$) Three-layer dispersion relations due to  (\ref{eq:SolDisp}) and two-layer dispersion relations due to  (\ref{eq:2layer1}) and (\ref{eq:2layer2}) of the Mg$||$Sb cell (table \ref{tab:Mg-Sb}) for different salt-layer heights $h_2 = 0.15,0.2,0.3,0.5$ and $5\, {\rm cm}$. The plus mode $\omega_+$ is red and the minus mode $\omega_-$ blue colored. The solid black line shows the two-layer frequency of the upper interface $\eta_1$ and the dashed black line the two-layer frequency of the lower interface $\eta_2$ both for $h_2 = 5\, {\rm cm}$. ($b$) Absolute value of the amplitude ratios due to  (\ref{eq:Ratio}) in dependence on $\epsilon_{mn}$ for $h_2 = 0.1,1.1,2.2,3.1,4.1,5.1,6.1$ and $7.1\, {\rm cm}$.}    % Bildunterschrift, optional
	\label{fig:Transition}          % Label für Verweise, optional
\end{figure}

\begin{figure}            % Einbetten in figure wie gehabt
	\centering                  % zentrierte Ausrichtung, optional
	\def\svgwidth{260pt}    
	% die Bildbreite muss auf diese Weise festgelegt werden!
	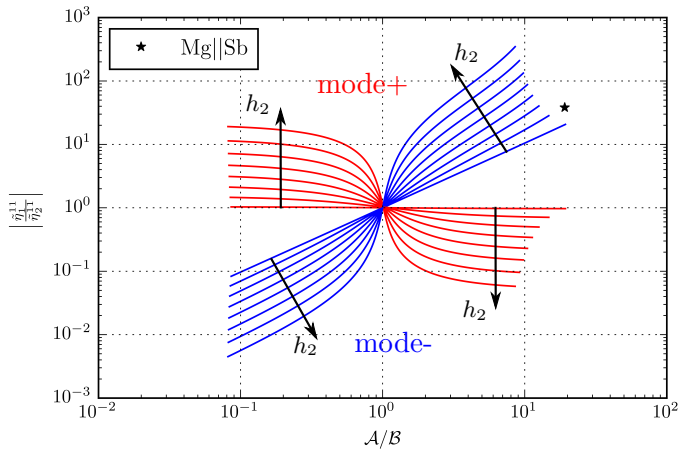  
	% 'versuchsaufbau' durch Dateinamen ersetzen.
	\caption{Absolute values of the amplitude ratios due to  (\ref{eq:Ratio}) for the metal pad roll mode $\epsilon_{11}$ in dependence of $\mathcal{A}/\mathcal{B}$
		with varying salt layer thicknesses $h_2 = 0.1,1,2,3,4,5,6$ and $7\, {\rm cm}$ exemplarily plotted using the generic Mg$||$Sb cell parameters from \cite{Weber2017} (table \ref{tab:Mg-Sb}). To diversify $\mathcal{A}$ and $\mathcal{B}$, $\rho_3$ was chanced from $1716\, \frac{{\rm kg}}{{\rm m^3}}$ up to $5000\, \frac{{\rm kg}}{{\rm m^3}}$.}    % Bildunterschrift, optional
	\label{fig:WaveRatio}          % Label für Verweise, optional
\end{figure}

In the following we analyze the strength of the interfacial wave coupling to predict in which parameter regimes the battery can be considered as decoupled and two-layer analysis to be sufficient to understand the MPRI, and in which regimes the wave coupling has a large share in the instability mechanism. \\
At fist we compare the expressions (\ref{eq:Omega12}$a$) and (\ref{eq:Omega23}$b$) of the three-layer dispersion relation with the corresponding two-layer dispersion relations $\omega_{\eta_{1}}$ and $\omega_{\eta_{2}}$ of both interfaces $\eta_1$ and $\eta_2$ that can be analogously derived and are given by 
\begin{subeqnarray}
	&&\omega_{\eta_{1}}^{2} = \frac{(\rho_2 - \rho_1)g\frac{\epsilon_{mn}}{R} + \gamma_{\eta_1}\left(\frac{\epsilon_{mn}}{R} \right)^3}{\rho_1\coth(\frac{\epsilon_{mn}}{R}h_1) + \rho_2\coth(\frac{\epsilon_{mn}}{R}h_2)}  \label{eq:2layer1} \\
	&&\omega_{\eta_{2}}^{2} = \frac{(\rho_3 - \rho_2)g\frac{\epsilon_{mn}}{R} + \gamma_{\eta_2}\left(\frac{\epsilon_{mn}}{R} \right)^3}{\rho_2\coth(\frac{\epsilon_{mn}}{R}h_2) + \rho_3\coth(\frac{\epsilon_{mn}}{R}h_3)}. \label{eq:2layer2}
\end{subeqnarray}
It becomes readily apparent that their deviation is only manifested in the terms
\begin{equation}
	\label{eq:dev}
	\frac{\tilde{\eta}_{2}^{mn}}{\tilde{\eta}_{1}^{mn}}\frac{1}{\sinh(\frac{\epsilon_{mn}}{R}h_2)} \ \ {\rm or} \ \  \frac{\tilde{\eta}_{1}^{mn}}{\tilde{\eta}_{2}^{mn}}\frac{1}{\sinh(\frac{\epsilon_{mn}}{R}h_2)},
\end{equation} 
respectively. These terms can be exploited to make first estimations in which cases the battery can be considered as decoupled. The terms are vanishing for two different limits: on the one hand if the salt layer height $h_2$ is becoming large in comparison to the radius $R$. In that limit both interfaces are placed too far apart from each other to interact and are both propagating in their own two-layer eigenfrequency $\omega_{\eta_{1}}$ and $\omega_{\eta_{2}}$ . Effectively, this limit is not practically relevant for LMBs since the salt layer should always be as thin as possible to realize the best possible efficiency. More important is on the other hand the second limit, where the amplitude ratio is becoming sufficiently small or large
\begin{equation}
	\frac{\tilde{\eta}_{1}^{mn}}{\tilde{\eta}_{2}^{mn}} \gg 1  \ \ {\rm or} \ \ \frac{\tilde{\eta}_{1}^{mn}}{\tilde{\eta}_{2}^{mn}} \ll 1. 
\end{equation}
In these limits one of the frequencies drops out and the other one reaches the two-layer limit. For which parameters this will be the case is principally determined by  (\ref{eq:Ratio}). However, in order to understand the key mechanisms of the interfacial coupling some deeper analyses are required. \\
In the first instance, to give the reader a better understanding, figure \ref{fig:Transition} shows exemplarily the frequencies ($a$) due to  (\ref{eq:SolDisp}) as well as the amplitude ratios ($b$) due to  (\ref{eq:Ratio}) for varying electrolyte layer heights in dependency of the wave number for the same Mg$||$Sb cell numerically 
investigated by \cite{Weber2017}. The used cell parameters and material properties are shown in table \ref{tab:Mg-Sb}. Figure \ref{fig:Transition}($a$) clarifies, how for increasing salt layer heights both frequency modes are progressively approaching the two-layer limits. But even more significant is the frequency transition appearing at $\epsilon_{mn} \approx 17$, where the two-layer frequencies are crossing each other. As emphasized in the zoomed window of Figure \ref{fig:Transition}($a$) showing the two layer frequencies $\omega_{\eta_{1}}$ and $\omega_{\eta_{2}}$ as well as the decoupled three-layer frequencies $\omega_{+}$ and $\omega_{-}$ for large salt layers, for small
wave numbers the fast mode $\omega_+$ is following the two-layer eigenfrequency $\omega_{\eta_{2}}$ of the upper interface and the slow mode $\omega_-$ is following the two-layer frequency $\omega_{\eta_{1}}$ of the lower interface, where from $\epsilon_{mn} \approx 17$ both modes behave in precisely the opposite manner. Hence, both modes cannot be uniquely assigned to the natural frequencies of uncoupled interfaces. Such a behavior has not been observed yet in coupled internal and surface waves \citep{Issenmann2016} and is leading to some new physical properties. Complementary, Figure \ref{fig:Transition}($b$) shows the absolute values of the amplitude ratio clarifying the strength of coupling in dependence on $\epsilon_{mn}$ and $h_2$. If we for instance focus on the smallest wave number $\epsilon_{11}$ representing the metal pad roll, one can see that for the Mg$||$Sb cell the amplitude ratios of both modes are not symmetric around $ |\tilde{\eta}_{1}^{11}/ \tilde{\eta}_{2}^{11}|$. If in that case the upper interface is excited, it will influence the lower one only very slightly. This is because the slow mode $\omega_-$ that is close to the eigenfrequency of the upper interface, will be excited and its amplitude ratio is very large here $ |\tilde{\eta}_{1}^{mn}/ \tilde{\eta}_{2}^{mn}|_- \approx 38$ such that $\tilde{\eta}_{2}^{mn}$ becomes negligibly small for reasonable values of $\tilde{\eta}_{1}^{mn}$. In contrast, perturbations of the lower interface with a eigenfrequency close to $\omega_+$ will substantially affect the upper one because there the amplitudes are strongly coupled $|\tilde{\eta}_{1}^{mn}/ \tilde{\eta}_{2}^{mn}|_+ \approx 0.7$. The above mentioned transition becomes also visible in the amplitude ratio lines at $\epsilon_{mn} \approx 17$ where all curves are crossing at $ |\tilde{\eta}_{1}^{mn}/ \tilde{\eta}_{2}^{mn}|_{\pm} \approx 1$ and both waves are perfectly coupled by both modes. That clarifies that not only the frequencies are turning, but also the manifestations of the modes within the interfaces. \\
\begin{table}
	\centering
	\begin{tabular}{lccc}
		Property & Layer 1 & Layer 2 & Layer 3 \\[4pt]
		$h$ (cm) &	4.5&				1&	4.5\\
		$R$ (cm) &	5&		5&	5\\	
		$\rho$ (g ${\rm cm}^{-3}$)	& 1.577 & 1.715 & 6.27 \\
		$\nu$ (${\rm m}^2$ ${\rm s}^{-1}$)	& $6.7\cdot 10^{-7}$ & $6.8\cdot 10^{-7}$ & $1.96\cdot 10^{-7}$\\
		\hline
		Property & Upper interface $\eta_1$ & Lower interface $\eta_2$\\
		$\gamma$ (N ${\rm m}^{-1}$)&0.19&0.095
	\end{tabular}
	\caption{Cell parameters of the Mg$||$Sb cell from \cite{Weber2017} used to calculate the wave frequencies and amplitude ratios of figure \ref{fig:Transition}.}
	\label{tab:Mg-Sb}
\end{table}
(\ref{eq:SolDisp}) together with (\ref{eq:Ratio}) are principally sufficient to completely describe the gravitational-capillary coupling dynamics. However, it is worthwhile also to predict analytically the transition point and to understand the physical nature behind this transition. While the presented transition as a function of the wave number is not of practical importance for LMBs, we have observed this transition also in dependence on many further cell parameters significantly changing for the various possible liquid metal and electrolyte combinations. We discovered that the influence of all cell parameters can be unified by only two dimensionless parameters.
We can proof that by eliminating the frequencies in  (\ref{eq:Omega12}$a$) and (\ref{eq:Omega23}$b$) what leads to a quadratic equation for the amplitude ratio:
\begin{eqnarray}
    \label{eq:AmplitudeRatio}
	&&a\left(\frac{\tilde{\eta}_{1}^{mn}}{\tilde{\eta}_{2}^{mn}}\right)^2 + b\frac{\tilde{\eta}_{1}^{mn}}{\tilde{\eta}_{2}^{mn}} - c = 0, \ {\rm with} \\
	&&a = \frac{\rho_2}{\sinh(\frac{\epsilon_{mn}}{R}h_2)\left(\rho_1\coth(\frac{\epsilon_{mn}}{R}h_1) + \rho_2\coth(\frac{\epsilon_{mn}}{R}h_2)\right)} \nonumber \\
	&&b = \left(\frac{(\rho_3 - \rho_2)g\frac{\epsilon_{mn}}{R} + \gamma_{\eta_2}\left(\frac{\epsilon_{mn}}{R} \right)^3}{(\rho_2 - \rho_1)g\frac{\epsilon_{mn}}{R} + \gamma_{\eta_1}\left(\frac{\epsilon_{mn}}{R} \right)^3} - \frac{\rho_2\coth(\frac{\epsilon_{mn}}{R}h_2) + \rho_3\coth(\frac{\epsilon_{mn}}{R}h_3)}{\rho_1\coth(\frac{\epsilon_{mn}}{R}h_1) + \rho_2\coth(\frac{\epsilon_{mn}}{R}h_2)}\right) \nonumber \\
	&&c = \frac{\rho_2}{\sinh(\frac{\epsilon_{mn}}{R}h_2)\left(\rho_1\coth(\frac{\epsilon_{mn}}{R}h_1) + \rho_2\coth(\frac{\epsilon_{mn}}{R}h_2)\right)}\cdot \frac{(\rho_3 - \rho_2)g\frac{\epsilon_{mn}}{R} + \gamma_{\eta_2}\left(\frac{\epsilon_{mn}}{R} \right)^3}{(\rho_2 - \rho_1)g\frac{\epsilon_{mn}}{R} + \gamma_{\eta_1}\left(\frac{\epsilon_{mn}}{R} \right)^3}. \nonumber 
\end{eqnarray}
As clarified in figure \ref{fig:Transition}($b$), the transition always appears at the intersection point of the modes, where the absolute values of the amplitude ratios have to equalize
\begin{equation}
	\left |\frac{\tilde{\eta}_{1}^{mn}}{\tilde{\eta}_{2}^{mn}}\right |_+  \stackrel{!}{=}
	\left |\frac{\tilde{\eta}_{1}^{mn}}{\tilde{\eta}_{2}^{mn}}\right |_-  .
\end{equation}
That is automatically fulfilled if (\ref{eq:AmplitudeRatio}) has only one solution, which in turn is guaranteed only if $b$ vanishes. 
This is again fulfilled only for
\begin{equation}
	\mathcal{A} \stackrel{!}{=} \mathcal{B},
\end{equation}
where the two dimensionless numbers $\mathcal{A}$ and $\mathcal{B}$ are defined with
\begin{eqnarray}
	&&\mathcal{A} := \frac{(\rho_3 - \rho_2)gR^2 + \gamma_{\eta_2}\epsilon_{mn}^{2}}{(\rho_2 - \rho_1)gR^2 + \gamma_{\eta_1}\epsilon_{mn}^{2}} \label{eq:A}\\
	&&\mathcal{B} := \frac{\rho_2\coth(\frac{\epsilon_{mn}}{R}h_2) + \rho_3\coth(\frac{\epsilon_{mn}}{R}h_3)}{\rho_1\coth(\frac{\epsilon_{mn}}{R}h_1) + \rho_2\coth(\frac{\epsilon_{mn}}{R}h_2)}. \label{eq:B}
\end{eqnarray}
Hence, the parameters $\mathcal{A}$ and $\mathcal{B}$ completely determine the transition. $\mathcal{A}$ may be interpreted as the ratio of the restoring gravity and capillary forces acting on both waves, whereas $\mathcal{B}$ describes the inertia distribution in the battery system. However, it can be shown that in practical cases the transition process is mainly captured by $\mathcal{A}$ alone. For the desired thin electrolyte-layer limit $h_2 \to 0$  (\ref{eq:AmplitudeRatio}) simplifies to 
\begin{equation}
	\left(\frac{\tilde{\eta}_{1}^{mn}}{\tilde{\eta}_{2}^{mn}}\right)^2 + (\mathcal{A} -1 )\frac{\tilde{\eta}_{1}^{mn}}{\tilde{\eta}_{2}^{mn}} - \mathcal{A} = 0.
\end{equation}
In this limit the amplitude ratio has become independent of $\mathcal{B}$ and offers two analytical solutions
\begin{equation}
	\frac{\tilde{\eta}_{1}^{mn}}{\tilde{\eta}_{2}^{mn}}_+ = 1  \ \ \ {\rm and} \ \ \
	\frac{\tilde{\eta}_{1}^{mn}}{\tilde{\eta}_{2}^{mn}}_- = -\mathcal{A}  \label{eq:ThinSolutions}
\end{equation}
From this we note that for thin salt layers the interfaces are perfectly coupled for all parameters when they are excited by the fast mode $\omega_+$, whereas the amplitude ratio of the slow mode $\omega_-$ is directly determined by $\mathcal{A}$.\\
Finally, $\mathcal{A}$ and $\mathcal{B}$ can be further exploited to physically understand the mode transition. At first we consider the limits of pure gravity and pure capillary waves. In these limits, $\mathcal{A}$ simplifies to 
\begin{equation}
	\mathcal{A}_{\rm gravity} = \frac{\rho_3 - \rho_2}{\rho_2 - \rho_1} \ \ \ {\rm or} \ \ \ \mathcal{A}_{\rm capillary} = \frac{\gamma_{\eta_2}}{\gamma_{\eta_1}}, \label{eq:ALimit}
\end{equation}
respectively. Hence, in large-scale LMBs the wave coupling is determined only by the ratio of the density differences. $\mathcal{A}_{\rm gravity}$ is the most important control parameter that should always be considered for predicting resonance. Further, the wave number-dependent mode transition, as shown in figure \ref{fig:Transition}, can be explained considering both limits: for the Mg$||$Sb cell we find $\mathcal{A}_{\rm gravity} > 1$ but $\mathcal{A}_{\rm capillary} < 1$. Consequently, this mode transition arose due to the change from gravity to capillary waves turning the sign of $\mathcal{A}$. \\
The coupling mechanism can be further understood by noting that $\mathcal{A}/\mathcal{B}$ can be expressed by the ratio of the two-layer frequencies
\begin{equation}
	\frac{\mathcal{A}}{\mathcal{B}} = \frac{\omega_{\eta_{2}}^{2}}{\omega_{\eta_{1}}^{2}}.
\end{equation}
Thus, as one would intuitively expect, the highest coupling indeed arises where the two-layer eigenfrequencies coincide and the interfaces step into resonance. The ratio $\mathcal{A}/\mathcal{B}$ quantifies the difference of the natural frequencies of both interfaces and therefore determines the strength of the coupling. For the important thin salt-layer limit $h_2 \to 0$, we further obtain
\begin{equation}
	\left |\frac{\tilde{\eta}_{1}^{mn}}{\tilde{\eta}_{2}^{mn}}\right |_- = \mathcal{A} = \frac{\omega_{\eta_{2}}^{2}}{\omega_{\eta_{1}}^{2}}, \label{eq:UpperLimit}
\end{equation}
showing that the amplitude ratio corresponding to the slow mode $\omega_-$ is directly determined by the ratio of the two-layer frequencies. \\
All these results are exemplified in figure \ref{fig:WaveRatio} showing the absolute values of the amplitude ratios in dependence of $\mathcal{A}/\mathcal{B}$ for different salt layer thicknesses of the Mg$||$Sb cell. All curves are indeed crossing at $\mathcal{A} = \mathcal{B}$. The two inner lines for the smallest chosen $h_2$ corresponds to the solutions (\ref{eq:ThinSolutions}) bounding the maximum coupling state. For large $\mathcal{A}/\mathcal{B} \gg 1$ the antisymmetric mode $\omega_-$ dominates in the upper interface $\eta_1$, while for low values $\mathcal{A}/\mathcal{B} \ll 1$ this mode is more prominent in the lower one $\eta_2$. The coupling state of the symmetric mode $\omega_+$ in contrast is not limited by $\mathcal{A}/\mathcal{B}$ since for sufficiently small $h_2$ the interfaces are always perfectly coupled by that mode.
\subsection{Coupling criterion for LMBs}
\label{sec:CouplingCriterion}
These findings now can be exploited to determine for which parameter regimes coupled dynamics must be considered and in which cases two-layer stability analysis, as practiced e.g. in the context of ARCs, is sufficient to understand interfacial instabilities in LMBs. At first it is desirable to predict which coupling mode will be excited by the arising Lorentz forces. As presented in the introduction, in two-layer systems the MPRI can be appropriately described by the Sele criterion. Since interfacial tension was respected for the coupling theory, it also has to be included into the Sele criterion. Following \cite{Gerbeau2006}, $\beta$ can be regarded as the ratio of the electromagnetic force due to interface displacements to the gravity force exerted on the interface due to the same displacements. By considering the sum of gravity and capillarity as the acting restoring force, we get modified Sele criteria for both interfaces
\begin{eqnarray}
&&\beta_{\eta_1} = \frac{J_0 b_z}{[(\rho_2 - \rho_1)gR^2 + \gamma_{\eta_1}\epsilon_{11}^{2}] h_1 h_2} > \beta_{\rm crit} \\ 
&&\beta_{\eta_2} = \frac{J_0 b_z}{[(\rho_3 - \rho_2)gR^2 + \gamma_{\eta_2}\epsilon_{11}^{2}] h_2 h_3} > \beta_{\rm crit}. 
\end{eqnarray}
From these criteria we can easily identify which interface is becoming unstable first. E.g., for large-size LMBs it is always the interface with the lower density difference. The ratio $\beta_{\eta_2}/\beta_{\eta_1}$ can be directly expressed in dependence of $\mathcal{A}$
\begin{equation}
\frac{\beta_{\eta_2}}{\beta_{\eta_1}} = \frac{((\rho_2 - \rho_1)gR^2 + \gamma_{\eta_1}\epsilon_{11}^{2})h_1}{((\rho_3 - \rho_2)gR^2 + \gamma_{\eta_2}\epsilon_{11}^{2})h_3} = \frac{1}{\mathcal{A}}\frac{h_1}{h_3} \label{eq:BetaRatio}
\end{equation} 
showing that for large $\mathcal{A}$ the upper interface will be more unstable while for small $\mathcal{A}$ the lower interface becomes more susceptible to electromagnetic excitation. By considering that at the first moment the natural two-layer frequencies will be excited until the third layer is interacting after some response time, it can be clarified that only the slow mode $\omega_-$ will be excited in LMBs for small and large values of $\mathcal{A}$. Because for large $\mathcal{A}$ the natural two-layer frequency of the first excited upper interface is slower and close to $\omega_-$ (see  (\ref{eq:Omega12}) or figure \ref{fig:Transition}($a$)) while for small $\mathcal{A}$ the lower interface becomes slower and even close to $\omega_-$. In both cases only the antisymmetric mode $\omega_-$ develops such that LMBs can principally decouple. On that basis a coupling criterion can be deduced by noting for the limit $h_2 \to 0$ the coupling is always on its highest level and neglecting $\omega_+$ we find from  (\ref{eq:ThinSolutions})
\begin{equation}
\left |\frac{\tilde{\eta}_{1}^{mn}}{\tilde{\eta}_{2}^{mn}}\right | \geq \mathcal{A} \  \ \ {\rm for} \ \ \mathcal{A} \gg 1  \ \ \ \ {\rm and} \ \ \ \ 
\left |\frac{\tilde{\eta}_{1}^{mn}}{\tilde{\eta}_{2}^{mn}}\right | \leq \mathcal{A} \ \ \ {\rm for} \ \ \mathcal{A} \ll 1. \label{eq:Decoupling}
\end{equation}
By defining a decoupling threshold such that the interfaces are assumed to be decoupled for $|\tilde{\eta}_{1}^{mn}/\tilde{\eta}_{2}^{mn}| > \chi$ or $|\tilde{\eta}_{1}^{mn}/\tilde{\eta}_{2}^{mn}| < 1/\chi$, we suggest the following criterion to evaluate the importance of wave coupling in LMBs: 
\begin{subeqnarray}
&&\eta_2 \approx 0 \ \ {\rm if} \left\{
\begin{array}{ll}
\mathcal{A} > \chi \frac{h_3}{h_1} &{\rm for} \ \ h_3 \geq h_1\\[2pt]
\mathcal{A} > \chi &{\rm for} \ \ h_3 < h_1 
\end{array} \right. \\
&&\eta_1 \approx 0 \ \ {\rm if} \left\{
\begin{array}{ll}
\mathcal{A} < \frac{1}{\chi}\frac{h_3}{h_1} &{\rm for} \ \ h_3 \leq h_1\\[2pt]
\mathcal{A} < \frac{1}{\chi} &{\rm for} \ \ h_3 > h_1 
\end{array} \right. \label{eq:CouplingCrit}
\end{subeqnarray}
These condition guaranty both that only one interface is excited due to  (\ref{eq:BetaRatio}) and the amplitude ratio is becoming large or small enough due to  (\ref{eq:Decoupling}). 
$\chi$ has to be chosen arbitrarily depending on the expediency. In the context of the metal pad roll dynamic we numerically found that $\chi = 10$ is suitable as will be discussed in the second part. In the case of the Mg$||$Sb battery numerically investigated with and without interfacial tension by \cite{Weber2017} and \cite{Bojarevics2017} we find high coupling parameters $\mathcal{A} \approx 28$ and $\mathcal{A}_{\rm gravity} \approx 33$ both clearly exceeding the coupling threshold. Indeed, both studies obtained wave motion essentially in the upper interface propagating with the frequency of the slow mode $\omega_-$. To determine the practical importance of wave coupling in large-size LMBs, $\mathcal{A}_{\rm gravity}$ was calculated for further metals. Table \ref{tab:A} shows the densities at operation temperature and the corresponding coupling parameter $\mathcal{A}_{\rm gravity}$ for most common metal and electrolyte combinations concerned as possible working materials for LMBs in literature. It can be seen that the so far studied Mg$||$Sb coincidentally typifies the extreme case of lowest wave coupling. Most batteries reveal values of $\mathcal{A}_{\rm gravity}$ between 3 and 6 where motion is present in both interfaces. Hence, coupling dynamics are indeed of high importance for most LMBs and may not be generally neglected as it has previously been assumed on the basis of the obtained dynamics in Mg$||$Sb cells. \\
Interestingly, the density ratios in three-layer aluminum-refining
cells are such that the interfaces in those cells are strongly coupled
as well (see table \ref{tab:A}). The following discussion applies therefore
equally to aluminum-refining cells. However, the refining process\textemdash
not limited by the open-circuit voltage\textemdash allows for much thicker
electrolyte layers, typically between 8 and $25\, {\rm cm}$
\citep{BeljajewRapoportFirsanowa:1957,PearsonPhillips:1957}.   
\begin{table}
	\centering
	\begin{tabular}{lcccccc}
		Electrodes & Electrolyte& $T$ ($^{\circ}{\rm C}$) & $\rho_1$ (kg $m^{-3}$) & $\rho_2$ (kg $m^{-3}$) & $\rho_3$ (kg $m^{-3}$)& $\mathcal{A}_{\rm gravity}$ \\[4pt]
		Li$||$Bi&	LiCl-LiF-LiI&485 &			488&	2690&	9800&									3.22\\
		Li$||$Pb&	LiCl-LiF-LiI&483	&			488&	2690&	10463&									3.53\\
		Li$||$Se&	LiCl-LiF-LiI&375	&			497&	2690&	3814&									0.51\\
		Li$||$Sn&	LiCl-LiF	&400	&		495&	1644&	6877&									4.55\\
		Li$||$Te&	LiCl-LiF-LiI&475	&			489&	2690&	5782&									1.41\\
		Li$||$Zn&	KCl-LiCl	&486	&		488&	1628&	6509&									4.28\\
		Mg$||$Sb&	KCl-MgCl$_2$-NaC&700	&			1577&	1715&	6270&									33.06\\
		Na$||$Bi&	NaCl-NaI-NaF&550	&			831&	2549&	9720&									4.18 \\
		Na$||$Pb&	NaCl-NaF-NaI&575	&			813&	2526&	9690&									4.18 \\
		Na$||$Sn&	NaCl-NaI&625	&			801&	2420&	6740&									2.67\\	        
		Ca$||$Bi&	CaCl$_2$-LiCl&550	&			1434&	1803&	9720&									21.43\\
		Ca$||$Sb&	CaCl$_2$-LiCl&700	&			1401&	1742&	6270&									13.28 \\
		K$||$Hg&	KBr-KI-KOH&250		&		640&	2400&	12992& 									6.02\\
		Al$||$Al-Cu$^{*}$ & AlF$_3$-NaF-CaCl$_2$-NaCl & 800 & 2300 & 2700 & 3140 & 	1.1	
	\end{tabular}
\caption{Coupling parameter $\mathcal{A}_{\rm gravity}$ calculated for different possible working material combinations. The densities are reported at working temperature $T$, see
\cite{Agruss1962a,Aqra2011,Bradwell2011,Bradwell2012,Cairns1967,Cairns1969b,Cairns1969c,Chum1980,Chum1981,Gale2004,Heredy1967,IAEA2008,Janz1976,Janz1979,Karas1963,Lyon1954,Shimotake1969,Sobolev2007,Sobolev2010,Spatocco2014,Swinkels1971,Wang2014,Weaver1962,Zinkle1998}. \mbox{}$^{*}$Aluminum-refining cell \citep{PearsonPhillips:1957} 
}
\label{tab:A}
\end{table}
\section{Numerical study of coupled interfacial waves}
All the above presented theoretical results and conclusions were derived only for inviscid gravity-capillary waves neglecting all electromagnetic effects. Consequently, it is necessary to verify the applicability of the Potential theory for more realistic electromagnetically driven battery systems. Further, from theory it is not evident which modes will evolve for strongly coupled interfaces $A \approx 1$ where both eigenfrequencies can be excited simultaneously. For this purpose, we conducted accompanying multiphase DNS in the predicted coupled-wave regime $0.1 < \mathcal{A} < 10$ taking into account the entire magnetohydrodynamics. We found three different coupling regimes with different stability properties. Besides the theoretically predicted antisymmetric motion for relatively small and large $\mathcal{A}$, novel propagation dynamics were discovered for strongly coupled interfaces $0.7 \lesssim \mathcal{A} \lesssim 2.1$ which will be presented in the following sections.
\subsection{Numerical set-up and procedure}
For the numerical study we used the three-dimensional multiphase DNS solver recently developed by \cite{Weber2017} particularly to investigate MHD instabilities in multilayer systems. The employed numerical model can be inspected in detail there.
On the basis of the extracted theoretical results a parameter study around $\mathcal{A} = 1$ was performed to investigate the wave dynamics in the regime of strongly coupled interfaces. To reach $\mathcal{A} = 1$, a generic standard case of equal metal layer heights $h_i$, conductivities $\sigma_i$, density differences $\Delta \rho$, interfacial tensions $\gamma_{\eta_1}$ and $\gamma_{\eta_2}$ as well as viscosities $\nu_i$ was defined. The parameters used are shown in table \ref{tab:B}. Because of the small height chosen for the electrolyte, the influence of $\mathcal{B}$ is very weak such that it is sufficient to describe the coupling only by $\mathcal{A}$. In order to capture also non-shallow water dynamics, all simulations were conducted in a aspect-ratio-one cylindrical tank of equal diameter $D$ and height $H$ of $10\, {\rm cm}$. The grid was constructed of purely orthogonal hexaedra with a lateral resolution of 50 cells and an axial resolution of 60 cells. This resolution was found to be sufficient to capture long-wave instabilities by employing a grid refinement study. The results did not significantly change when using finer grids. \\
Various simulation series were conducted for different values of $\mathcal{A}$. $\mathcal{A}$ was modified stepwise by adapting only the density of the upper layer $\rho_1$. As predicted by the theory, we achieved the same results by modifying other $\mathcal{A}$-dependent parameters. For all single parameters $\mathcal{A}$ various simulation with different cell currents $I_0$ were performed to find the stability onsets by successively increasing $I_0$. Once the current becomes large enough ($I_0 > I_c$) to destabilize the system, small numerical errors will grow and form the metal pad roll after a while. 
After first noticeable interfacial displacements were identified for $I_0 > I_c$, we ensured to leave the system sufficient time to  reach a steady wave propagation state or to develop a short-circuit.  Sufficient means here that we only accepted wave propagation as converged if both, the interface amplitudes and the mean velocity in the cell, were found to be constant for at least 30 periods. By that procedure 
the propagation regime between the rest state and the short-circuit was identified for all investigated $\mathcal{A}$ in dependence of the cell current.
Finally, exemplary simulations in the propagation regime were chosen to evaluate the wave frequencies and amplitude ratios. It was ensured that the amplitudes were sufficiently small to allow comparison with linear wave theory.
\begin{table}
	\centering
	\begin{tabular}{lccc}
		Property & Layer 1 & Layer 2 & Layer 3 \\[4pt]
		$h$ (cm) &	4.5&				1&	4.5\\	
		$\rho$ (g ${\rm cm}^{-3}$)	& 2.5 & 3 & 3.5 \\
		$\sigma$ (S ${\rm m}^{-1}$)	& $10^6$ & 500 & $10^6$ \\
		$\nu$ (${\rm m}^2$ ${\rm s}^{-1}$)	& $6.7\cdot 10^{-7}$ & $6.7\cdot 10^{-7}$ & $6.7\cdot 10^{-7}$\\
		$b_z$ (mT)	& 10 & 10 & 10 \\
		\hline
		Property & Upper interface $\eta_1$ & Lower interface $\eta_2$\\
		$\gamma$ (N ${\rm m}^{-1}$)&0.1&0.1
	\end{tabular}
	\caption{Used generic simulation parameters of the standard case yielding $\mathcal{A} = 1$.}
	\label{tab:B}
\end{table}
\subsection{Numerical results}
For all investigated coupling parameters interfacial motion could be confirmed in accordance with the presented theoretical predictions. In general, we found two distinguishable coupling regimes in dependence on $\mathcal{A}$ comprising fundamentally different coupling dynamics. The first regime was found for $0.7 \lesssim \mathcal{A} \lesssim 2$, where the amplitudes of both interfaces are similarly strong. This regime involves complex dynamics and various possible wave modes that can evolve. We call it from now on the "strongly coupled regime". Respectively, for $\mathcal{A}\lesssim 0.7$ and $\mathcal{A} \gtrsim 2$ a second regime, later on referred to as the "weakly coupled regime", was found, where always one interface is dominant. In these simulations we indeed obtained only the antisymmetric coupling $\omega_-$ as predicted by the theory. At first, characteristic modes observed for both regimes are introduced qualitatively in the following section. For reasons of clarity, the weakly coupled regime is presented first.
\subsubsection{Weakly coupled regime} 
\label{sec:Weakly}
\begin{figure}            % Einbetten in figure wie gehabt
	\centering                  % zentrierte Ausrichtung, optional
	\def\svgwidth{380pt}    
	% die Bildbreite muss auf diese Weise festgelegt werden!
	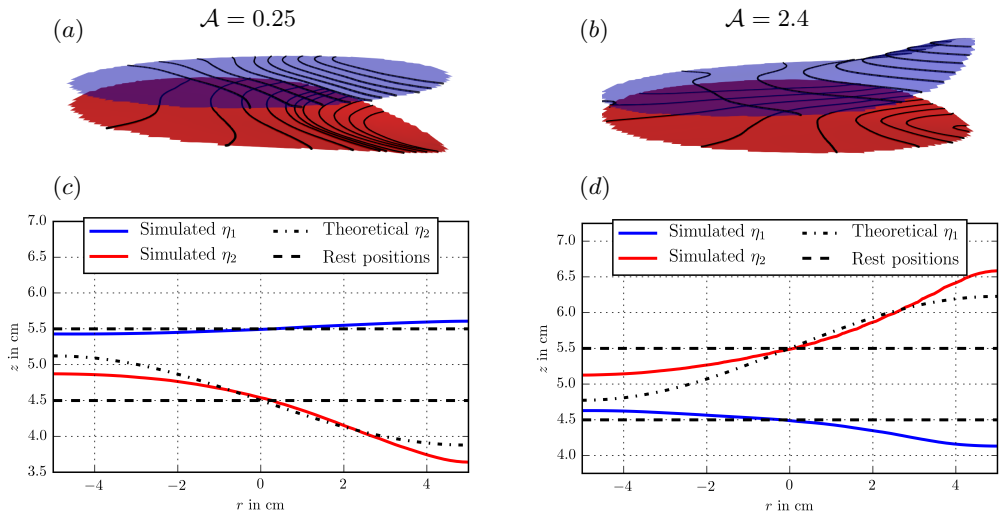  
	% 'versuchsaufbau' durch Dateinamen ersetzen.
	\caption{Exemplary contour plots of the excited interfaces for coupling parameters $\mathcal{A} = 0.25$ ($a$) and $\mathcal{A} = 2.4$ ($b$). ($c$) and ($d$) cross sections of the corresponding interface contours $\eta_1$ (blue) and $\eta_2$ (red) in the $r$-$z$ planes orientated along the highest amplitude shaping. In addition, the dashed lines mark the rest position of the interfaces and the dotted lines show the radial component of the theoretical gravity-capillary modes due to (\ref{eq:eta1}) and (\ref{eq:eta2}). }    % Bildunterschrift, optional
	\label{fig:ContourModes}          % Label für Verweise, optional
\end{figure}

First of all the obtained interfacial shapes and directions of rotation are qualitatively analyzed.
Figures \ref{fig:ContourModes}($a$) and \ref{fig:ContourModes}($b$) show the contours of both interfaces for the exemplarily chosen values $\mathcal{A} = 0.25$ and $\mathcal{A} = 2.4$, respectively. It can be seen that in both cases the interfaces are anti-symmetrically displaced, where for $\mathcal{A} = 0.25$ the lower and for $\mathcal{A} = 2.4$ the upper interface is more strongly deformed. The resulting amplitude ratios can be identified more clearly in figures \ref{fig:ContourModes}($c$) and \ref{fig:ContourModes}($d$), showing cross sections of the interface contours in the $r$-$z$ planes orientated along the highest amplitude shaping. It becomes apparent that the amplitude of the subordinate upper interface for $\mathcal{A} = 0.25$ is significantly smaller than the amplitude of the subordinate lower interface for $\mathcal{A} = 2.4$. That is in qualitative agreement with the theoretical slow mode amplitude ratios yielding disparate values of $|\tilde{\eta}_{1}^{mn}/\tilde{\eta}_{2}^{mn}|_- = 0.18$ and $|\tilde{\eta}_{1}^{mn}/\tilde{\eta}_{2}^{mn}|_- = 2.8$. In addition, figures \ref{fig:ContourModes}($c$) and \ref{fig:ContourModes}($d$) further show the theoretical radial shape due to (\ref{eq:eta1}$a$) and (\ref{eq:eta2}$b$) of the superordinate interfaces. Clear deviations between the numerical and analytical gravitational solutions can be recognized. In comparison to the gravitational modes, the magnetically driven interfacial waves are not symmetrically displaced around the rest positions. Rather, the interfaces tip away from each other such that the emergence of a short-circuit is counteracted. Hence, the theory tends here to overestimate the threat of short-circuits. \\
Further, for both exemplary cases different directions of rotations of the interfaces were observed. For $\mathcal{A} = 0.26$ both interfaces rotate synchronously clockwise and for $\mathcal{A} = 2.4$ counterclockwise. The same behavior holds true for all investigated $\mathcal{A}\lesssim 0.7$ (clockwise) and  $\mathcal{A} \gtrsim 2$ (counterclockwise). Hence, the direction of rotation in the weakly coupled regime is uniquely determined by $\mathcal{A}$ (and also $\mathcal{B}$ for unpractical high electrolyte layers) and the initial question from the introduction concerning the unknown rotation direction is clarified. This behavior can be explained by the fact that the larger displaced interface, that is determined by $\mathcal{A}$, causes higher horizontal compensation currents and thus a larger resulting Lorentz force than the smaller one. Since we know from the coupling analysis that both interfaces can only rotate synchronously in the same direction, the overall coupled rotation is determined by the Lorentz force associated to the larger displaced interface that overcompensates the oppositely directed Lorentz force caused by the small interface.
\subsubsection{Strongly coupled regime}
\label{sec:HighlyCoupledRegime}
\begin{figure}            % Einbetten in figure wie gehabt
	\centering                  % zentrierte Ausrichtung, optional
	\def\svgwidth{380pt}    
	% die Bildbreite muss auf diese Weise festgelegt werden!
	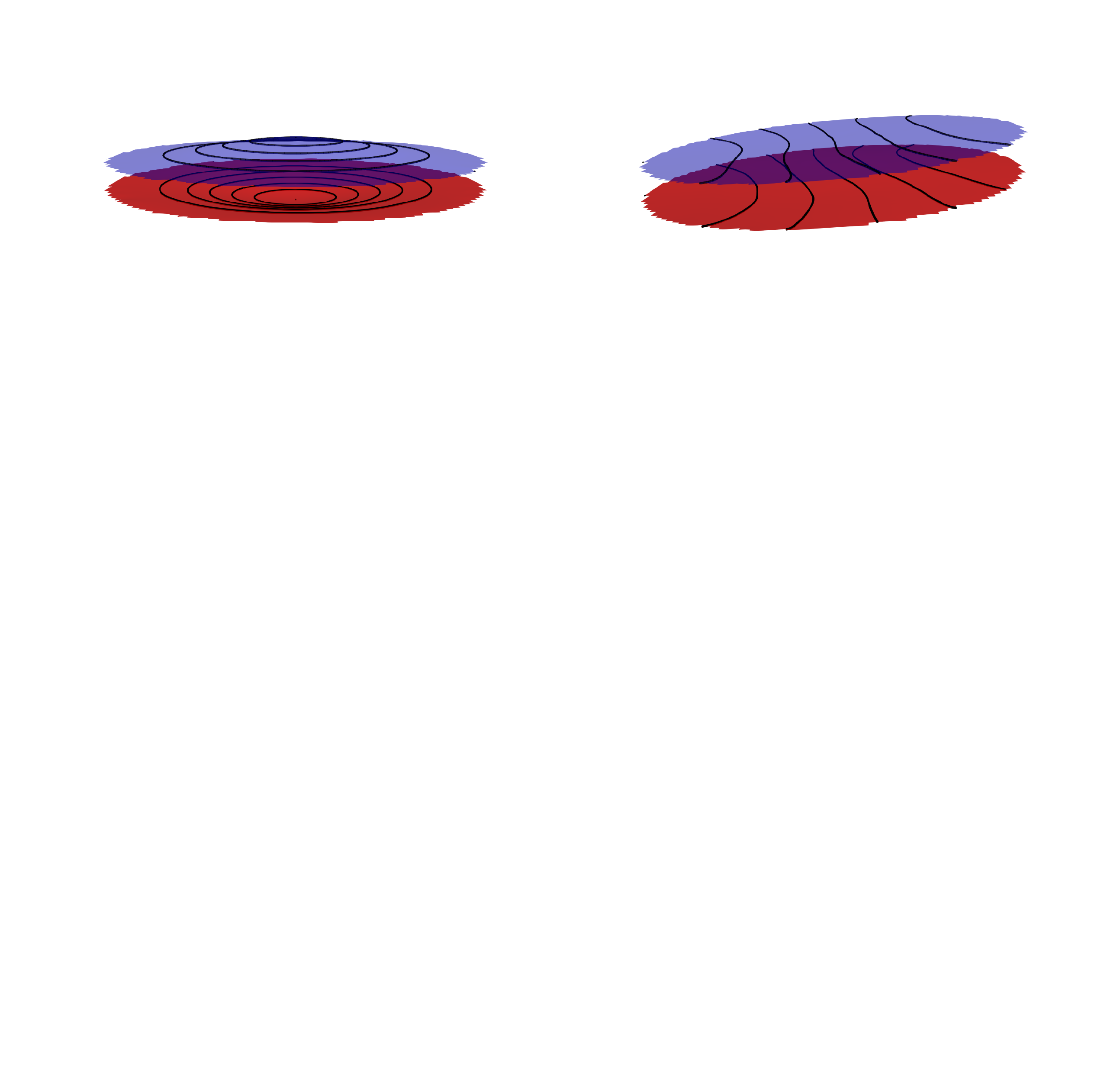  
	% 'versuchsaufbau' durch Dateinamen ersetzen.
	\caption{Exemplary contour plots of the excited interfaces for the coupling parameter $\mathcal{A} = 1.1$ with the applied cell currents of $I = 250\, {\rm A}$ ($a$) and $I = 500\, {\rm A}$ ($b$). These currents lead to antisymmetric and symmetric wave-coupling, respectively. ($c$) and ($d$) cross sections of the corresponding interface contours $\eta_1$ (blue) and $\eta_2$ (red) in the $r$-$z$ planes orientated along the highest amplitude shaping. (c) shows in addition the fitted curves due to (\ref{eq:Shape1}) and (\ref{eq:Shape2}). ($e$) and ($f$) depict further the induced compensation current vectors in the same cross section. A superimposed color map highlights the radial (horizontal) component of the compensation currents.}    % Bildunterschrift, optional
	\label{fig:ContourNewInst}          % Label für Verweise, optional
\end{figure}
For coupling parameters closer to one ($0.7 \lesssim \mathcal{A} \lesssim 2$) far more complex dynamics were observed. Note first that these regime thresholds are purely empirical at this stage. They may change in dependence on $h_2$ (affecting $\mathcal{B}$) and possibly on further parameters not occurring in $\mathcal{A}$ and $\mathcal{B}$ like the viscosities and electrical conductivities. In this regime, we found different coupling modes and excited wave numbers as well as a non-oscillatory instability in dependence of the cell current. Further, time-dependent mode transitions were found such that one mode can merge into another one spontaneously making analysis more complicated. However, two characteristic flow states were found that are mainly present in this regime exemplarily shown in figure \ref{fig:ContourNewInst}. For relatively low applied cell currents slightly above the observed stability thresholds, we always found oppositely-directed radial-symmetric rotating interfaces, as exemplarily shown in figure \ref{fig:ContourNewInst}($a$) and figure \ref{fig:ContourNewInst}($c$). This state largely differs from the MPRI in the physical sense since we observed that these vortex shapes can persist for minutes or even for the complete simulation depending on the cell currents. Hence, it can be considered as a quasi time-independent non-oscillatory instability that can not be described by the wall reflection of oscillating interfacial waves, compare to section \ref{sec:Introduction}.
Figure \ref{fig:ContourNewInst}($e$) shows in addition the induced radial compensation currents in cross section. In contrast to the metal pad roll mode, two closed current loops can be identified leading to pure azimuthal Lorentz forces inducing opposite directed circular motions in the liquid electrodes. In section \ref{sec:CentripitalInst} we give a detailed explanation how centripetal pressure differences can cause this instability.\\
For higher applied cell currents, a second characteristic state was found to develop: symmetrically coupled in-phase rotating waves with the expected wavenumber $\epsilon_{11}$ associated to the metal pad roll, as depicted in figures \ref{fig:ContourNewInst}($b$) and \ref{fig:ContourNewInst}($d$). Also for high cell currents close to the short-circuit onset, the radial instability arose first. But for sufficiently high cell currents this coupling state always spontaneously turned to the symmetric mode after some time, then remaining preserved, or\textemdash in only very few cases\textemdash once again turning to the previously presented anti-symmetric mode of the weakly coupled regime (sec. \ref{sec:Weakly}). This instability even strongly differs from the common MPRI mechanism and seems to be counterintuitive at first glance since the metal pad is always explained by a redistribution of the cell currents due to changes of the electrical resistivity caused by electrolyte-layer deformations. But for (almost) perfect parallel interfaces, as presented here, there is no preferred direction for the currents to chose because the salt-layer height remains constant everywhere. Thus, horizontal compensation currents are not expected to arise. However, it can be seen from figure \ref{fig:ContourNewInst}($f$), showing the radial compensation currents in the cross section orientated along the highest amplitude shaping, that here horizontal currents appear in the electrolyte. This is in contrast to the common assumption stating that the current remains almost vertical in the low-conducting electrolyte. A detailed explanation unraveling that issue is given in section \ref{sec:Symm}. \\
For the sake of completeness, it should be noted that besides the modes presented here and the antisymmetric mode of the weakly coupled regime we sometimes observed also higher modes or mode superpositions in particular for high cell currents near the short-circuit onset. However, these waves were only temporarily stable and seemed to appear randomly. All in all, the two modes presented here were highly dominant and can be considered as the characteristic modes of the strongly coupled regime, though the mode dynamics was found to be far more complex and can not be precisely predicted at this stage.
\subsection{Comparison with the Potential theory}
\begin{figure}            % Einbetten in figure wie gehabt
	\centering 
	\hspace*{-0.25cm}                 % zentrierte Ausrichtung, optional
	\def\svgwidth{390pt}    
	% die Bildbreite muss auf diese Weise festgelegt werden!
	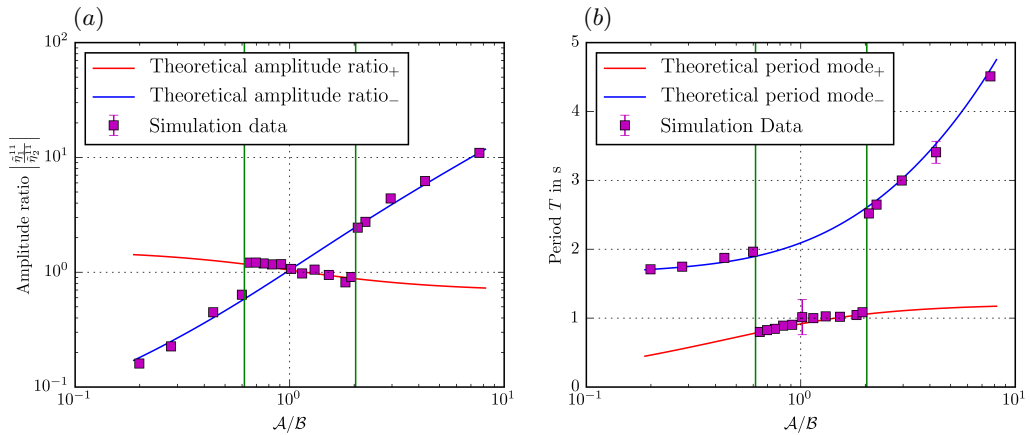  
	% 'versuchsaufbau' durch Dateinamen ersetzen.
	\caption{Numerical and analytical amplitude ratios ($a$) and wave periods ($b$) in dependence of the coupling parameter $\mathcal{A}/\mathcal{B}$. The strongly coupled regime is bounded by two vertical lines. For clarity, only the most present symmetric mode $\omega_+$ is shown in the strongly coupled regime.}    % Bildunterschrift, optional
	\label{fig:Comparison}          % Label für Verweise, optional
\end{figure}
Both the amplitudes ratios and the wave frequencies are quantitatively analyzed and compared with the formulas (\ref{eq:SolDisp}) and (\ref{eq:Ratio}). To calculate the numerical amplitude ratios, we tracked the minimum and maximum displacement of both interfaces at every time-step. Since the interfaces are not necessarily evenly displaced around the initial positions, as shown in figures \ref{fig:ContourModes}($c$) and \ref{fig:ContourModes}($d$), the amplitude ratios are calculated by the differences of the maximum and minimum displacements
\begin{equation}
\frac{\tilde{\eta}_{1}}{\tilde{\eta}_{2}} = \frac{\max(\eta_1) -\min(\eta_1)}{\max(\eta_2) -\min(\eta_2)}.
\end{equation}
In order to reduce uncertainties, $\tilde{\eta}_{1}/\tilde{\eta}_{2}$ was evaluated for a wide range of time-steps comprising at least 50 periods and averaged. The wave periods were calculated by tracking the angle of highest position of the interfaces over time and fitting the angular velocity. Figures \ref{fig:Comparison}($a$) and \ref{fig:Comparison}($b$) show the amplitude ratios determined in such a way and wave periods for all evaluated $\mathcal{A}/\mathcal{B}$ in comparison with the Potential curves. For clarity, we only show the dominant symmetric mode $\omega_+$ in the strongly coupled regime marked by two vertical lines. Both the numerical amplitude ratios as well as the wave periods are in agreement with the theory for both modes. Consequently, though the interfacial shapes are considerably deviating from the Potential solutions, it is confirmed that\textemdash at least in very good approximation\textemdash  eigenmodes of pure gravity-capillary waves are excited by the induced Lorentz forces. Further it can be concluded that the wave coupling indeed is mainly determined by pressure exchanges. Contributions of induced current redistributions to the coupling as well as viscous effects seem to be of little relevance. Hence, the highly simplified (and often underestimated) irrotational Potential theory is sufficient to describe coupled interfaces driven by a rotational Lorentz force.    
\section{Analysis and interpretation of the coupled instabilities}
In this section we aim to give first possible explanations of the two instabilities in the strongly coupled regime and analyze how they affect the battery. From now on, we denote the instability causing the radial-symmetric state from figure \ref{fig:ContourNewInst}$a$ as the \textit{bulge instability} (BI) due to its memorable shape. The symmetric rolling motion presented in figure \ref{fig:ContourNewInst}$b$ we call in the following the \textit{synchronous tilting instability} (STI) in accordance to the characteristic synchronous motion.
\subsection{Bulge instability}
\label{sec:CentripitalInst}
\begin{figure}            % Einbetten in figure wie gehabt
	\centering                  % zentrierte Ausrichtung, optional
	\def\svgwidth{150pt}    
	% die Bildbreite muss auf diese Weise festgelegt werden!
	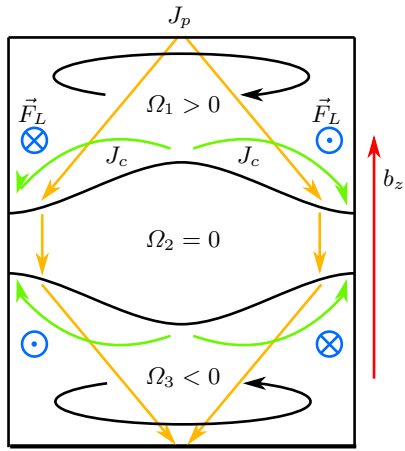  
	% 'versuchsaufbau' durch Dateinamen ersetzen.
	\caption{Schematic vertical cut of the BI. The current $\vec{J}_p$ is redirected to the tank wall giving rise to horizontal compensation currents $\vec{J}_c$ which, in interaction with an external vertical magnetic field $b_z$, in turn lead to azimuthal Lorentz forces $\vec{F}_L$ inducing counter-rotating vortices with angular velocities $\Omega_1$ and $\Omega_3$. The concave and convex interface shapes then arise to equilibrate centripetal pressure differences.}
	\label{fig:RadialInst}          % Label für Verweise, optional
\end{figure}

The BI, that was found only in the strongly coupled regime, can be regarded as a special instability case since it appeared as a quasi-steady state that may persist over minutes. Hence, it can not be described by time-dependent gravity-capillary waves. Rather, it should be considered as a further fixed point existing in the system. In fact, stationary interface perturbations also have been observed before in physically similar ARCs by \cite{Bojarevics2006} and were later quantitatively described by \cite{Munger2008b} for cylindrical geometries. They identified differences between the centripetal pressure of both phases that have to be balanced by interface deformations, as the key mechanism. This phenomenon is widely present in everyday life and can be seen, e.g., when stirring a cup of tea. Figure \ref{fig:RadialInst} illustrates the driving mechanism due to \cite{Munger2008b} transferred to three-layer LMBs. At first, the explanation is the same as for the MPRI.
Radial-symmetrically displaced interfaces will lead to also radial-symmetric compensation currents $\vec{J}_c$ since the perturbed current $\vec{J}_p$ chooses the path near the circular tank walls where the electrolyte is thin. In interaction with $b_z$, these currents lead to pure azimuthal Lorentz forces $\vec{F}_L$ driving two large-scale vortices spinning clockwise in the anode- and counterclockwise in the cathode-layer. The vortices can be described by a priori unknown average radial angular velocities profiles $\Omega_1 (r)$ and $\Omega_2 (r)$, whereas no average rotation is expected to arise in the electrolyte-layer $\Omega_2 \approx 0$. Thereby, centripetal pressure jumps arise at the interfaces reading
\begin{eqnarray}
&&\Delta P|_{\eta_1} = \frac{1}{2}\rho_1 \Omega_{1}^{2}r^2 > 0 \\
&&\Delta P|_{\eta_2} = -\frac{1}{2}\rho_3 \Omega_{3}^{2}r^2 < 0. 
\end{eqnarray}
These pressure drops can be equilibrated only by a concave deformation of $\eta_1$ locally decreasing the hydrostatic pressure $P_g$ and a convex deformation of $\eta_2$ increasing $P_g$, respectively. To provide a more quantitative comprehension that the BI is caused by centripetal pressure drops, we further studied the interfacial shapes. Since it is very complicated to calculate Lorentz force-induced angular velocities profiles, we assumed that the Lorentz force is acting like a mechanical stirrer. For unbaffled stirred tanks, the angular velocity can be described by \citep{Platzer1983}
\begin{equation}
\Omega (r) = \Omega_0 \left[\frac{1}{2} + \frac{1}{2}\left(\frac{r}{r_0}\right)^4   \right]^{-\frac{1}{2}},
\end{equation} 
where $\Omega_0$ and $r_0$ denote the rotation frequency and characteristic length of the impeller. Applied to the metal layers, this profile leads to the theoretical vortex shapes
\begin{eqnarray}
&&\eta_{1}(r) = h_1 - \tilde{\eta}_{1} - \frac{(\Omega_{0,1} r_{0,1})^2}{g}\arctan\left(\frac{r}{r_{0,1}} \right)^2 \label{eq:Shape1} \\
&&\eta_{2}(r) = h_3 - \tilde{\eta}_{2} + \frac{(\Omega_{0,2} r_{0,2})^2}{g}\arctan\left(\frac{r}{r_{0,2}} \right)^2 .
\label{eq:Shape2}
\end{eqnarray} 
The characteristic frequencies and length-scales $\Omega_{0,1}$, $\Omega_{0,2}$, $r_{0,1}$ and $r_{0,2}$ corresponding to the Lorentz forces are not known. Therefore, the curves (\ref{eq:Shape1}) and (\ref{eq:Shape2}) were fitted to the simulated interfaces exemplarily shown in figure \ref{fig:ContourNewInst}($c$). The fits are in perfect agreement with the interfaces, also in the other simulations, which implies that the centripetal pressure approach is in principal suitable to describe the radial symmetric ground state observed in the LMB model. However, further modeling is mandatory and planned for future studies to describe the magnetically induced angular velocity profiles and amplitudes in dependence of the cell current. The analogy to mechanical stirring presented here should be considered just as a first approach to identify the key mechanism that very likely causes these shapes. 
For ARCs \cite{Munger2008b} found that non-oscillatory instabilities evolve half as fast as the MPRI, which is why it was rarely found in simulations. However, in three-layer LMBs strongly coupled interfaces favor this instability, because the opposite rotating liquid metals produce stronger shearing forces in the electrolyte causing higher pressure drops at the interfaces in comparison to ARCs, where the aluminum and the cryolite rotate in the same direction with different angular velocities. Hence, the centripetal pressure differences are even stronger such that the growth rate of the BI can apparently exceed the growth of the MPRI, when both interfaces are simultaneously excited. Either way since the chosen cylindrical geometry naturally favors the forming of radial-symmetric vortex shapes, we do not expect the BI being that significant in rectangular containers.
\subsection{Synchronous tilting instability}
\label{sec:Symm}
\begin{figure}            % Einbetten in figure wie gehabt
	\centering                  % zentrierte Ausrichtung, optional
	\def\svgwidth{320pt}    
	% die Bildbreite muss auf diese Weise festgelegt werden!
	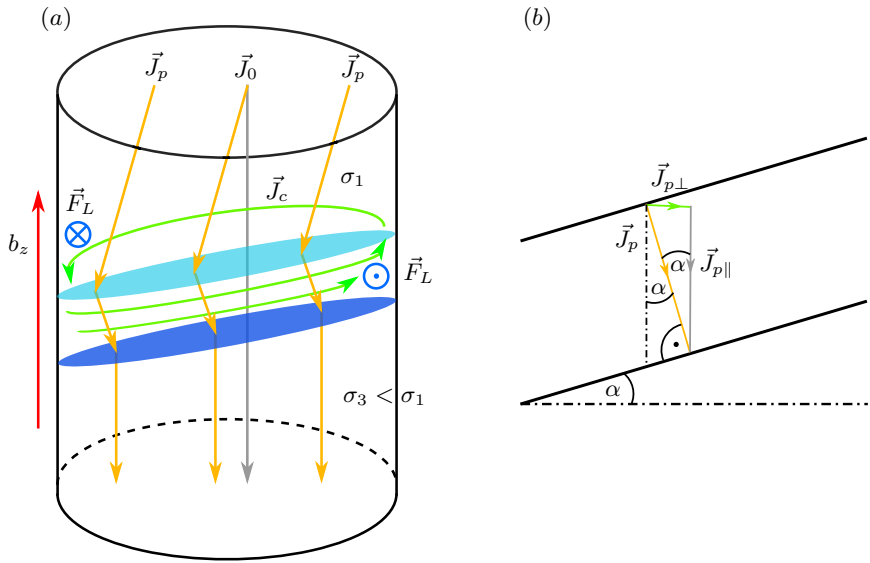  
	% 'versuchsaufbau' durch Dateinamen ersetzen.
	\caption{Schematic explanation of the STI. ($a$) profile of an LMB showing schematically the paths of the unperturbed cell current $\vec{J}_0$, the perturbation current $\vec{J}_p$ caused by the symmetrically tilted interfaces and the resulting horizontal compensation current $\vec{J}_c$. $\vec{J}_c$ closes here in the upper layer of higher electrical conductivity $\sigma_1 > \sigma_3$. The interaction of $\vec{J}_c$ with an external vertical magnetic field $b_z$ leads to azimuthal Lorentz forces mainly in the electrolyte driving the rotating waves. ($b$) electrolyte layer section. The perturbation current $\vec{J}_p$ in the electrolyte can be decomposed into a component $\vec{J}_{p\parallel}$ parallel to $b_z$ and a component $\vec{J}_{p\perp}$ vertical to $b_z$. The ratio of both components is connected by a pitch angle $\alpha$ to the slope of the interfaces.}
	\label{fig:SymmetricInst}          % Label für Verweise, optional
\end{figure}

As already mentioned in section \ref{sec:HighlyCoupledRegime}, also the discovered STI can not be explained by the Sele mechanism alone (see section \ref{sec:Introduction}) since the electrolyte-layer height remains (almost) preserved such that no preferred direction for the current is expected to develop. We observed that this mechanism is considerably stabilizing in the manner that higher cell currents are necessary to provoke short-circuits in comparison to the antisymmetric coupling of the weakly coupled regime. Hence, it is highly desirable to understand the physical mechanism behind this instability that can deliver practical benefits for LMB operation. Figure \ref{fig:SymmetricInst}($a$) clarifies schematically the driving mechanism we assume behind the STI. The key assumption is, again, that the current always takes the shortest path through the electrolyte due to its orders higher electrical resistivity. When the interfaces are displaced in parallel, the perturbed current $\vec{J}_p$ must pass perpendicularly through the electrolyte, as sketched in figure \ref{fig:SymmetricInst}($a$). This current has, in dependence on the electrolyte layer's inclination, a horizontal component leading to Lorentz forces acting within the electrolyte and driving the rotation of both interfaces simultaneously. Further, the horizontal component that can be regarded as acting like the compensation current $\vec{J_c}$ of the classical mechanism described in figure \ref{fig:Sele}, has to close either in the anode or cathode layer (as long the electrical conductivities of the liquid metals are larger than of the current collectors \citep{Munger2006b}. It will close in the better conducting metal as shown in figure \ref{fig:SymmetricInst}($a$) for the case $\sigma_1 > \sigma_3$. Here, the current in the anode layer is deflected to the left in order to minimize the current path through the more resistive cathode layer. Additional simulations with varying electrical conductivities, not shown here, confirmed this explanation. For equal electrode layer conductivities, as used in the coupling study, the stronger displaced interface and thus $\mathcal{A}$ determined the closing layer. \\
Compared to the MPRI, the STI allows much higher currents without leading to a short-circuit. We will give here a possible explanation by comparing the induced horizontal currents of both instabilities.
The horizontal currents arising in the STI can be quantitatively estimated by decomposing the perpendicular perturbation current $\vec{J_p}$ into a component $\vec{J}_{p\parallel}$ parallel to the magnetic field $b_z$, which does not contribute to the Lorentz force, and a component $\vec{J}_{p\perp}$ vertical to $b_z$. $\vec{J}_{p\perp}$ is acting like the horizontal compensation current of the Sele-mechanism $\vec{J}_{p\perp} \hat{=} \vec{J}_c$ and drives the instability. Figure \ref{fig:SymmetricInst}($b$) shows these decomposition applied to a small electrolyte layer section appearing approximately linear. By assuming that $\vec{J}_p$ is equally distributed in the electrolyte layer, we identify $|\vec{J}_p| \approx |\vec{J}_0|$ within since the three layers can be considered as a serial connection. From figure \ref{fig:SymmetricInst}($b$) we find the relation
\begin{equation}
\sin(\alpha) = \frac{J_{p\perp}}{J_0} 
\end{equation}
that can be further equalized with the interfacial slope. For the purposes of a conservative estimate the interfaces are approximated only by the radial  component $\eta_r$ in direction of the strongest interfacial displacement. From  (\ref{eq:Surf1}) the radial slope of both interfaces is given by
\begin{equation}
\frac{\partial \eta_r (r)}{\partial r} = \tilde{\eta}\frac{\epsilon_{11}}{R}\frac{1}{2}\left( J_0(\epsilon_{11}\frac{r}{R}) -J_2(\epsilon_{11}\frac{r}{R})\right)
\end{equation}
leading to the $r$-dependent horizontal perturbation current
\begin{equation}
|\vec{J}_{p\perp}| = |\vec{J}_0|\tilde{\eta}\frac{\epsilon_{11}}{R}\frac{1}{2}\left( J_0(\epsilon_{11}\frac{r}{R}) -J_2(\epsilon_{11}\frac{r}{R})\right).
\end{equation}
In order to estimate the global impact to the interfaces, both the maximum horizontal current $\max|\vec{J}_{p\perp}|$ at $r=0$ as well as the the mean current $\langle |\vec{J}_{p\perp}|\rangle_r$ are evaluated
\begin{eqnarray}
&&\max|\vec{J}_{p\perp}| = |\vec{J}_0|\frac{\epsilon_{11}}{2}\frac{\tilde{\eta}}{R} \label{eq:Horicurrent} \\
&&\langle |\vec{J}_{p\perp}|\rangle_r = |\vec{J}_0|J_1 (\epsilon_{11})\frac{\tilde{\eta}}{R}. \label{eq:MeanCurrent}
\end{eqnarray}
The maximum current $\max|\vec{J}_{p\perp}|$ can be easily exploited for comparisons with the maximum compensations currents obtained in the simulation. For the case shown in figure \ref{fig:ContourNewInst}($f$) we find for example an average amplitude of $\tilde{\eta} \approx 0.85\, {\rm cm}$ with the applied current density of $|\vec{J}_0| = 6.37\, {\rm A}/{\rm cm^2}$ and $R=5\, {\rm cm}$. These values lead to maximum horizontal currents of $\max|\vec{J}_{p\perp}| \approx 1\, {\rm A}/{\rm cm^2}$ matching well with the highest compensation currents occurring in figure \ref{fig:ContourNewInst}($f$). Good agreement was found as well for the other simulations within the strongly coupled regime, giving an promising indication that indeed the simple vector decomposition is sufficient to approximate horizontal compensation currents caused by the symmetric wave mode.  \\
Finally, we analyze the potential of the STI to delay the short-circuit in comparison to the MPRI present in the weakly- or decoupled regimes. Since it is unclear how the wave coupling of the antisymmetric mode in the weakly coupled regime exactly effects the stability, the classical Sele-mechanism describing the MPRI in ARCs is used for comparisons. As known from literature \citep{Gerbeau2006}, in these systems the induced horizontal compensation currents $|\vec{J}_{c}|_{{\rm ARC}}$ in the aluminum layer are of the order
\begin{equation}
|\vec{J}_{c}|_{{\rm ARC}} \sim \frac{|\vec{J}_0| \tilde{\eta}_1 2R}{h_1 h_2}. \label{eq:SeleCurrent}
\end{equation}
Transferred to the LMBs, this expression is a valid estimation for $\mathcal{A}/\mathcal{B} \gg 1$ (the estimation can be done just as well for $\mathcal{A}/\mathcal{B} \ll 1$), where only the upper interface is excited by the slow mode $|\vec{J}_{c}|_{{\rm ARC}} \hat{=} |\vec{J}_{c}|_{{\rm LMB,}\omega_-}$. The ratio of $|\vec{J}_{c}|_{{\rm LMB,}\omega_-}$ to the mean horizontal current produced by the STI (\ref{eq:MeanCurrent}) that are induced by the same interface displacements, scales with 
\begin{equation}
\frac{|\vec{J}_{c}|_{{\rm LMB,}\omega_-}}{|\vec{J}_{c}|_{{\rm LMB,}\omega_+}} \sim \frac{R^2}{h_1 h_2},
\end{equation}
Since the horizontal currents drive both instabilities, we can conclude that the MPRI will lead faster to short-circuits as the STI because we find $h_1 , h_2 < R$ in practice. This effect will be even more prominent in shallow batteries, as they are desired for practical operation in order to minimize Ohmic losses. Hence, this stabilizing mechanism can be exploited to potentially improve the operation safety of LMBs, eg., by modifying the density of the electrolytes to approach the strongly coupled regime. However, we provide only a first understanding of this novel instability mechanism. Quantitative three-layer stability analysis and further numerical studies, which are necessary to gain a deeper understanding, are left for future work.  
\section{Concluding remarks} 
In this paper we investigated how coupled interfaces in liquid metal batteries affect the interfacial wave dynamics arising due to the metal pad roll instability. Potential analysis was performed to investigate the pressure coupling dynamics of interfacial gravity-capillary waves in a battery model of cylindrical shape. An analytical dispersion relation (\ref{eq:SolDisp}) as well as equations for the amplitude ratio (\ref{eq:Ratio}) and (\ref{eq:AmplitudeRatio}) were derived which lead to two different coupling modes. Both interfaces can propagate either symmetrically in phase or antisymmetrically phase-shifted by $180^{\circ}$. Further, we extracted two independent dimensionless parameters $\mathcal{A}$ (\ref{eq:A}) and $\mathcal{B}$ (\ref{eq:B}) completely determining the overall coupling dynamics, where for shallow electrolytes only one coupling parameter $\mathcal{A}$ is sufficient to predict the coupling behavior. Using these results, we developed a decoupling criterion (\ref{eq:CouplingCrit}) that was applied to different possible working metals. It was shown that the interfacial wave coupling will be indeed present in most future LMBs (table \ref{tab:A}) and may not be generally neglected, only for some particular working materials like the previously examined Mg$||$Sb cell. \\
Accompanying multiphase direct numerical simulations were conducted to verify the applicability of the Potential theory to viscous interfacial waves driven by the magnetohydrodynamical metal pad roll instability. Various simulations were performed to analyze the impact of the coupling strength in dependence of $\mathcal{A}$ to the system dynamics and stability. The numerical wave frequencies and amplitude ratios are in very good agreement to the theory (figures \ref{fig:Comparison}$a$ and \ref{fig:Comparison}$b$) confirming that the interfacial coupling strength in liquid metal batteries can be well approximated by the Potential theory. 
Merely the interfacial shapes deviate visibly from the pure gravity-capillary modes. \\
Beside that, three different coupling regimes involving different characteristic coupling modes were numerically identified in dependence on $\mathcal{A}$. Firstly, the decoupled regime, where only one interface is excited and the overall instability mechanism is very close to two-layer aluminum reduction cells. Secondly, we defined a weakly coupled regime where both interfaces are excited and always antisymmetrically coupled (figure \ref{fig:ContourModes}). And finally, we observed more complicated dynamics for strongly coupled interfaces, called the strongly coupled regime, respectively. Here, essentially two different and novel coupling states can be found (figure \ref{fig:ContourNewInst}). A radial-symmetric rotational instability for low cell currents and symmetrically coupled metal pads for higher cell currents. Both states can not be described by the classical metal pad roll mechanism alone. We identified differences of the centripetal pressure for the radial instability (figure \ref{fig:RadialInst}) and horizontal currents occurring in the electrolyte (figure \ref{fig:SymmetricInst}) for the symmetric instability as the essential driving mechanisms and provided first quantitative justifications. The symmetric instability was found to have less potential for short-circuiting the cell compared to the metal pad roll instability such that future LMBs may potentially benefit from hydrodynamically coupled interfaces. Hence, the investigation of instabilities associated to the strongly coupled regime appears very worthwhile and promising for further research. 

\section*{Acknowledgements}
This  work  was  supported  by  Helmholtz-Gemeinschaft  Deutscher  Forschungszentren (HGF) in frame of the Helmholtz Alliance “Liquid metal technologies” (LIMTECH). Fruitful  discussions  with  Frank Stefani, Oleg Zikanov, Caroline Nore and Wietze Herreman on several aspects of metal pad roll instability and liquid metal batteries are gratefully acknowledged.

\bibliographystyle{jfm}
\bibliography{literature}

\end{document}